\documentclass[12pt, a4paper]{article}
            
\usepackage{a4}
\usepackage{graphicx}
\usepackage{amsmath}
\usepackage{amssymb}
\usepackage{subfigure}
\usepackage[all]{xy}  
\usepackage{color}
\usepackage{mathtools}

\usepackage{hyperref}
\hypersetup{
			pdftitle={ Landscape Study of  Target Space Duality of (0,2) Heterotic String Models    }, 
      pdfauthor={Ralph Blumenhagen, Thorsten Rahn}, 
      pdfsubject={Algebraic Geometry, Mathematical Methods in Physics},
      pdfstartview={FitH}, pdfpagelayout={TwoColumnRight},
      bookmarksopen, bookmarksnumbered, bookmarksopenlevel=2}

\definecolor{darkblue}{rgb}{0,0,.5}
\definecolor{lightblue}{rgb}{0.8,0.85,1}
\definecolor{mygray}{gray}{.75}

\newcommand{\cohomCalgKoszul}{{\text{\fontfamily{put}\bfseries\footnotesize\selectfont cohomCalg Koszul} extension}}
\newcommand{\cohomCalg}{{\text{\fontfamily{put}\bfseries\footnotesize\selectfont cohomCalg }}}

\newcommand{\eq}[1]{\begin{equation}
                     \begin{split} #1 \end{split}
                     \end{equation}}
\newcommand{\beq}{\begin{equation}}  \newcommand{\eeq}{\end{equation}}
\newcommand{\bal}{\begin{aligned}}   \newcommand{\eal}{\end{aligned}}

\def\IP{\mathbb{P}}

\def\cM{\mathcal{M}}
\def\cO{\mathcal{O}}

\def\clap#1{\hbox to 0pt{\hss#1\hss}}
\def\mllap{\mathpalette\mathllapinternal}

\def\mathllapinternal#1#2{%
\llap{$\mathsurround=0pt#1{#2}$}}

\def\fto{\longrightarrow}
\def\injto{\lhook\joinrel\relbar\!\!\:\joinrel\rightarrow}
\def\surjto{\relbar\joinrel\twoheadrightarrow}

\def\uspc{${}^\big.$}   

\begin{document}

\baselineskip=14pt

\vspace*{-1.5cm}
\begin{flushright}    
  {\small  MPP-2011-71}
\end{flushright}

\vspace{2cm}
\begin{center}        
  {\LARGE
  Landscape Study of  Target Space Duality of  \\[0.4cm] 
   $(0,2)$ Heterotic String Models    
  }
\end{center}

\vspace{0.75cm}
\begin{center}        
  Ralph Blumenhagen and  Thorsten Rahn
\end{center}

\vspace{0.15cm}
\begin{center}        
  \emph{Max-Planck-Institut f\"ur Physik, F\"ohringer Ring 6, \\ 
               80805 M\"unchen, Germany} \\[5mm]
\end{center} 

\vspace{2cm}


\begin{abstract}
In the framework of $(0,2)$ gauged linear sigma models, we systematically
generate sets of perturbatively dual heterotic string compactifications. 
This target space  duality is first derived in non-geometric phases
and then translated to the level of GLSMs and its geometric phases.
In a landscape analysis, we compare  the massless
chiral spectra and the dimensions of the moduli spaces.
Our study includes geometries given by complete intersections
of hypersurfaces in toric varieties equipped with 
$SU(n)$ vector bundles defined via the monad construction.
\end{abstract}

\vspace{1.5cm}

\begin{center}
{\small {\bf Dedicated to the memory of Maximilian Kreuzer}}
\end{center}
\clearpage

\newpage


\tableofcontents

\newpage
\section{Introduction}

String compactifications to four dimensions with ${\cal N}=1$
supersymmetry have been under intense study since 
the mid eighties. Various constructions have been considered,
which include heterotic strings on Calabi-Yau threefolds, 
type II orientifolds on Calabi-Yau threefolds with
intersecting D-branes or more recently F-theory 
on singular elliptically fibered fourfolds\footnote{M-theory
compactifications on singular $G_2$ manifolds are certainly
the most poorly understood models.}.
Models of this kind constitute the most promising way
to relate string theory to the real world and therefore
have also been studied from various phenomenological points
of view. 

From a string theory perspective it is also important to
better understand the moduli spaces of these models.
Due the low amount of supersymmetry this question is rather involved,
as there can be a D-term and an  F-term potential. These,  
on the one hand, provide the potential to destabilize the model
or on the other hand, with sufficient ingredients, lead to a stabilization of 
the moduli at finite values. In addition, it is one of the
salient features of string theory that seemingly different 
geometric backgrounds can lead to the same string theory,
i.e.~conformal field theory on the world-sheet.
Well known examples include $T$-duality for toroidal
compactifications or mirror symmetry for type II compactifications
on Calabi-Yau threefolds. 
The redundancy provided by such perturbative (in $g_s$) dualities
is both a beautiful mathematical aspect of string theory respectively
quantum geometry and a structure which has to be taken into account
in a landscape study of string compactifications.

In this paper, we  study such target space dualities in the
context of heterotic string compactifications defined
via the $(0,2)$ gauged linear sigma model (GLSM) \cite{Witten:1993yc}.
The GLSM provides an overall description of the complexified
compact stringy K\"ahler moduli space, which is divided 
into various cones (phases) \cite{Aspinwall:1993nu}, which can be either geometric or non-geometric.
In the geometric phases, the GLSM is equivalent  to a non-linear
sigma model with a Calabi-Yau target space ${\cal M}$ with additional
left-moving world-sheet fermions coupling to the connection of a holomorphic
(stable) vector bundle ${\cal V}$ on ${\cal M}$. 
For a non-standard embedding, i.e.~${\cal V}\ne T_{\cal M}$, these
are the heterotic $(0,2)$  models first considered in 
\cite{WittenCohomology,Distler:1987ee}. 
For ${\cal V}=T_{\cal M}$ 
the world-sheet supersymmetry
enhances to $(2,2)$, and mirror symmetry gives a pair of dual models
$({\cal M},T_{\cal M})\simeq ({\cal W},T_{\cal W})$. The question is whether also two
seemingly completely different geometric configurations
$({\cal M},{\cal V})$ and $(\widetilde{\cal M},\widetilde{\cal V})$ can be target space dual.

That such a duality might exist was first pointed out in \cite{DistlerKachruDuality},
where it was observed that in the non-geometric phase, which
in the simplest case is given by a $(0,2)$ Landau-Ginzburg (LG)  model,
two models can be trivially equivalent, as they give rise
to the same Landau-Ginzburg superpotential. The nature of this
duality in one of the phases of the GLSMs could have essentially
be of two different kinds: a.) since often the LG model has
more singlets than at generic points in the geometric phases,
it could be that there is a transition between two $(0,2)$ models.
Hence, it would be like a conifold transition between two Calabi-Yau threefolds
in type II; \ b.) the two $(0,2)$ models could be isomorphic (target
space dual), which just happens to be directly visible in the LG phases.
This situation is rather  like mirror symmetry.

As was mentioned in \cite{DistlerGreeneFeatures} 
and explicitly checked for a few
simple examples in \cite{RalphTargetSpace1,RalphTargetSpace2}, 
evidence for the second possibility
could be provided by computing the total number of massless
gauge singlets  of  candidate dual pairs in the two geometric phases. 
Geometrically this number corresponds to the dimension of
the space of first order  deformations of the two models
and is subdivided into K\"ahler, complex structure and
bundle deformations. 
Thus, a necessary condition for the existence of a target space
duality (T-duality) between  two  $(0,2)$ models in their geometric phases
is
\begin{equation}
     h^{1,1}({\cal M}) + h^{2,1}({\cal M}) +  h^{1}_{\cal M}({\rm
       End}({\cal V})) =
     h^{1,1}(\widetilde{\cal M}) + h^{2,1}(\widetilde{\cal M}) +  h^{1}_{\widetilde{\cal M}}({\rm
       End}(\widetilde{\cal V})) \; .\nonumber
\end{equation}
This sum of dimensions of cohomology classes 
is not necessarily the dimension of the global
moduli space, as there can be obstructions among  the complex
structure and bundle moduli, captured in the effective 
four-dimensional supergravity theory by a non-trivial superpotential.
Therefore, these sums  are not constant over the moduli space, i.e.
there exist subloci where they jump. The proposed  target space  duality
would mean that such jumps are also mapped consistently.   
It is well known that the superpotential can receive contributions 
only at  sigma model tree-level
and from  non-perturbative world-sheet instantons.  
In addition, the chiral matter spectra should also be identical, 
i.e.
\eq{h^{i}_{\cal M}(\wedge^k\, {\cal V})  = h^{i}_{\widetilde{\cal M}}(\wedge^k\,
\widetilde{\cal V}),\qquad  {\rm for}\  \ i=0,\ldots, 3 \nonumber
}
with $k=0, \ldots, n$  depending on the rank of
the $SU(n)$ bundles. 

The objection of this paper is two-fold: First, we want to study,
how with the help of the $(0,2)$ GLSM potentially dual $(0,2)$
models can be generated. We will see that the existence of 
a  LG-phase is not necessary and that the existence of
any non-geometric phase suffices to provide candidate dual
models.  Second, with the  recently released \cohomCalg\
and \cohomCalgKoszul\
packages \cite{Blumenhagen:2010pv,Blumenhagen:2010ed,cohomCalg:Implementation} 
we have now the means to compute the
relevant dimensions of cohomology classes, in particular
the bundle moduli space $h^{1}({\cal M},{\rm End}({\cal V}))$
\cite{Distler:1988jj} in an automatic
manner\footnote{The quite tedious computations in
  \cite{RalphTargetSpace1,RalphTargetSpace2} were carried out by hand.}
so that we can essentially perform  a sort of $(0,2)$ landscape study of
the above cohomological identity for dual models.
Thus, our study will involve dual classes whose models are quite generic complete intersection
Calabi-Yau threefolds equipped with quite generic $SU(n)$  monad bundles.

Let us mention that for these monad constructions the proof
of $\mu$-stability is notoriously difficult so that for our study
we will just assume that they are stable. We understand
that in certain cases, stability might not be satisfied, but,
since we are doing a landscape study, we are certain that this
will not affect our general conclusion. In this respect, it would be interesting
to generalize the results of \cite{Anderson:2007nc,Anderson:2008uw,ExploringPositiveMonads} to vector bundles over generic
toric varieties, which are not products of projective spaces.

The paper is organized as follows: In section 2 we give a brief introduction 
into $(0,2)$ gauged linear sigma models and how their 
non-geometric Landau-Ginzburg phase motivates a kind of
target space duality between two seemingly different GLSMs. We also 
discuss how such $(0,2)$ GLSMs define GUT like four-dimensional
compactifications of the $E_8\times E_8$ 
heterotic string and how in geometric phases the massless matter modes 
are related to various vector bundle valued cohomology groups.
 
In section 3 we present how the target space duality first established
in the LG-phase, can be generalized to other hybrid-type
non-geometric phases. This should be considered 
as a quite generic algorithm  for   the
determination of possible target space dual $(0,2)$ GLSMs. 
Then, we discuss a couple of examples, for which we compare
the massless spectra in the geometric phases of  dual pairs. 
Moreover, we point out that geometrically
the Calabi-Yau base manifolds of dual pairs are in certain cases related via 
conifold transitions. These provide examples of the so-called
transgressions  of vector bundles as introduced in \cite{Candelas:2007ac}.

Section 4 is devoted to our report on a landscape study of many thousands
of dual models, which are based on the lists of Calabi-Yau manifolds
defined via hypersurfaces and complete intersections of two hypersurfaces
in toric ambient spaces. Our results  provide  compelling  evidence
for the existence of target space-dualities for heterotic string
compactifications with ${\cal N}=1$ space-time supersymmetry in 
four dimensions. We emphasize that this goes way beyond a
$(0,2)$ generalization of mirror symmetry, which would just be
a $\mathbb Z_2$ symmetry, whereas here one can generate many
dual $(0,2)$ models.

\section{Basic ingredients}

In this section we review a couple of well known facts on the
$(0,2)$ gauged linear sigma model and its application
for heterotic model building with ${\cal N}=1$ space-time
supersymmetry and GUT gauge groups. 

\subsection[Basics  of (0,2) gauged linear sigma models]{Basics  of \boldmath{$(0,2)$} gauged linear sigma models}
\label{sec_GLSM}

The framework we are working in is the $(0,2)$ GLSM introduced
in \cite{Witten:1993yc}. Let us briefly review a couple
of important aspect. For a more thorough introduction we refer to
the original literature.

The GLSM is a  massive two-dimensional field theory 
which is 
believed, under suitable conditions, to flow in the infrared to a non-trivial 
superconformal field theory. Moreover, it is closely related to toric
geometry. 
The classical vacua of the GLSM do depend on the values of the
Fayet-Iliopoulos terms leading  to a cone structure, which
captures the cone structure of the complexified K\"ahler moduli
space of Calabi-Yau compactifications.   
These so-called  phases torically correspond
to the various triangulations of a polytope resulting in a collection of cones in a fan \cite{Aspinwall:1993nu}.
At low energies, these 
phases appear to correspond to theories such as a non-linear  sigma-model, a 
Landau-Ginzburg orbifold, or some other more peculiar theory like a hybrid 
model. 

More concretely, let us first list the fields in the $(0,2)$ GLSM.
We only consider abelian gauge symmetries so that we have
a number of $U(1)$ gauge fields $A^{(\alpha)}$ with $\alpha=1,\ldots, r$. 
There 
are two sets of chiral superfields:  $\{X_i\vert i=1,\ldots, d\}$ with 
$U(1)^r$ charges
$Q^{(\alpha)}_i$ and $\{ P_l\vert l=1,\ldots,\gamma \}$ with $U(1)$ charges 
$-M^{(\alpha)}_l$. To eventually describe compact Calabi-Yau manifolds, 
we assume that $Q^{(\alpha)}_i\ge 0$ and that for each $i$, there exist at least
one $r$ such that  $Q^{(\alpha)}_i> 0$.
Furthermore, there are two sets of Fermi superfields: $\{\Lambda^a\vert 
a=1,\ldots, \delta\}$ 
with charges $N^{(\alpha)}_a$ and $\{ \Gamma^{(\alpha)}_j\vert j=1,\ldots,c\}$ with 
charges $-S^{(\alpha)}_j$. We also assume that the charges $M^{(\alpha)}_l$,
$N^{(\alpha)}_a$ and $S^{(\alpha)}_j$ satisfy the same (semi-)positivity constraints
as the $Q^{(\alpha)}_i$.
In the following we specify such a GLSM by writing all the above
data in a table of the form
\begin{equation}
\begin{aligned}
&\begin{array}{|c||c|}
\hline
x_i & \Gamma^j \\
\noalign{\hrule height 1pt}
\begin{array}{cccc}
 Q^{(1)}_1 & Q^{(1)}_2 & \ldots\ldots & Q^{(1)}_d\\
 Q^{(2)}_1 & Q^{(2)}_2 & \ldots\ldots & Q^{(2)}_d\\
 \vdots  &    \vdots  &  \vdots  &  \vdots  \\  
 Q^{(r)}_1 & Q^{(r)}_2 & \ldots\ldots & Q^{(r)}_d
\end{array}
&
\begin{array}{cccc}
-S^{(1)}_1 & -S^{(1)}_2 & \ldots\ldots & S^{(1)}_c\\
-S^{(2)}_1 & -S^{(2)}_2 & \ldots\ldots & S^{(2)}_c\\
 \vdots  &    \vdots  &  \vdots  &  \vdots  \\  
-S^{(r)}_1 & -S^{(r)}_2 & \ldots\ldots & S^{(r)}_c\\
\end{array}\\
\hline
\end{array}
\\[0.1cm]
&\begin{array}{|c||c|}
\hline
\Lambda^a & p_l \\
\noalign{\hrule height 1pt}
\begin{array}{cccc}
 N^{(1)}_1 & N^{(1)}_2 & \ldots\ldots & N^{(1)}_\delta\\
 N^{(2)}_1 & N^{(2)}_2 & \ldots\ldots & N^{(2)}_\delta\\
 \vdots  &    \vdots  &  \vdots  &  \vdots  \\  
N^{(r)}_1 & N^{(r)}_2 & \ldots\ldots & N^{(r)}_\delta
\end{array}
&
\begin{array}{cccc}
-M^{(1)}_1 & -M^{(1)}_2 & \ldots\ldots & -M^{(1)}_\gamma\\
-M^{(2)}_1 & -M^{(2)}_2 & \ldots\ldots & -M^{(2)}_\gamma\\
 \vdots  &    \vdots  &  \vdots  &  \vdots  \\ 
-M^{(r)}_1 & -M^{(r)}_2 & \ldots\ldots & -M^{(r)}_\gamma\\
\end{array}\\
\hline
\end{array}\,.
\end{aligned}
\end{equation}
where the index $\alpha=1,\ldots, r$ will be suppressed in most cases.
Gauge and gravitational anomaly cancellation of the  two-dimensional
GLSM requires the 
following set of quadratic and linear constraints to be satisfied
\begin{equation}\label{eq_anomcancel}
\begin{aligned}
  \sum_{a=1}^\delta  N_a^{(\alpha)} = \sum_{l=1}^\gamma  M_l^{(\alpha)},\qquad\quad
   &\sum_{i=1}^d  Q_i^{(\alpha)} = \sum_{j=1}^c  S_j^{(\alpha)} \\ 
  \sum_{l=1}^\gamma  M_l^{(\alpha)} M_l^{(\beta)} -   \sum_{a=1}^\delta  N_a^{(\alpha)} N_a^{(\beta)} 
  =&  \sum_{j=1}^c  S_j^{(\alpha)}  S_j^{(\beta)}  - \sum_{i=1}^d Q_i^{(\alpha)} Q_i^{(\beta)} \; ,
\end{aligned}
\end{equation}
for all $\alpha, \beta=1,\ldots, r$. 

Besides the chiral and Fermi superfields, a GLSM is defined via
a non-trivial superpotential of the form
\begin{equation}
\label{ftermsuperpot}
 S=\int d^2 z d\theta\,  \left[ \sum_j \Gamma^j\,  G_j(X_i) + 
               \sum_{l,a} P_l\,  \Lambda^a\,  F_a{}^l(X_i)  \right], 
\end{equation}
where $G_j$ and $F_a{}^l$ are quasi-homogeneous polynomials whose multi-degree is 
fixed by  requiring charge neutrality of the action. 
Moreover, they satisfy the transversality constraint that
$F_a{}^l(X)=0$ only for $X_i=0$. 
The multi-degrees of the polynomials $G_j$ and $F_a{}^l$ are given
in the following table
\begin{equation}
\begin{aligned}
&\begin{array}{|c|}
\hline
 G^j \\
\noalign{\hrule height 1pt}
\begin{array}{cccc}
 S_1 & S_2 & \ldots\ldots & S_c
\end{array}\\
\hline
\end{array}
\\[0.1cm]
&\begin{array}{|c|}
\hline
F_a{}^l  \\
\noalign{\hrule height 1pt}
\begin{array}{cccc}
 M_1-N_1 & M_1-N_2 & \ldots\ldots & M_1-N_\delta\\
 M_2-N_1 & M_2-N_2 & \ldots\ldots & M_2-N_\delta\\
 \vdots  &    \vdots  &  \vdots  &  \vdots  \\  
M_\gamma-N_1 & M_\gamma-N_2 & \ldots\ldots & M_\gamma-N_\delta
\end{array}\\
\hline
\end{array}\,.
\end{aligned}
\end{equation}

\noindent
In addition to the induced  F-term scalar potential 
\begin{equation}
  V_F= \sum_j \Bigl\vert G_j(x_i)\Bigr\vert^2  + 
               \sum_a \Bigl\vert \sum_{l} p_l\,   F_a{}^l(x_i)  \Bigr|^2 
\end{equation}
there also appears a D-term scalar potential. Introducing the
Fayet-Iliopoulos parameter $\xi^{(\alpha)}\in \mathbb R$ for each $U(1)$ it simply reads
\begin{equation}
  V_D= \sum_{\alpha=1}^r   \biggl(  \sum_{i=1}^ d  Q_i^{(\alpha)} \vert x_i \vert^2 - \sum_{l=1}^\gamma
     M_l^{(\alpha)} \vert p_l \vert^2 -\xi^{(\alpha)} \biggr)^2 
\end{equation}
where $x_i$ and $p_l$ are the bosonic complex scalars of  the
corresponding chiral superfields. 

For a concrete choice of charges one can now determine the classical vacua
of the F-term and D-term potential. It turns out that the structure
of this vacuum depends crucially on the Fayet-Iliopoulos terms.
In fact the $\mathbb R^r$ parametrized by them splits into cones, also
called phases, whose boundaries separate different vacuum configurations.
Let us briefly discuss this for the most simple choice 
of a single $U(1)$ and $\gamma=1$. In this case there is only
a single Fayet-Iliopoulos parameter and one only obtains
two different phases:

For $\xi> 0$   the D-term implies that not all $x_i$ are
allowed to vanish simultaneously. Thus not all $F_a$ do vanish
and vanishing of the F-term potentials implies $G_j(x_i)=0$ and
$\langle p\rangle=0$.
Thus in this phase one gets   a $(0,2)$  non-linear sigma-model 
on a generally singular complete intersection in a weighted projective space,
$\IP_{Q_1,\ldots,Q_{d}}[S_1,\ldots,S_{c}]$. Moreover, the
superpotential \eqref{ftermsuperpot} induces for  
the fermionic components  $\lambda_a$ of  the Fermi superfields $\Lambda^a$ 
the mass term
\begin{equation}
     L_{\rm mass} = \sum_a \pi\, \lambda^a F_a \; ,
\end{equation}
which, due to the transversality condition, means
that one linear combination  of the $\lambda^a$ receives a mass
by pairing up with the fermionic component $\pi$ of the
chiral superfield $P$.
More generally, each $\pi_l$ pairs up with a linear combination
of the $\lambda_a$ so that the massless combinations
of the left-moving fermions $\lambda_a$ couple to 
a coherent sheaf ${\cal V}$  of rank $\text{rk}({\cal
  V})=\delta-\gamma-r_{\cal V}$
defined as the cohomology of the monad
\begin{equation}\label{eq_general monad}
 0\rightarrow \cO_{\cM}^{\oplus r_{\cal V}}\;  {\buildrel \otimes E_i{}^a \over
   \longrightarrow}\;  \bigoplus_{a=1}^{\delta} \cO_\cM(N_a) \; {\buildrel \otimes F_a{}^l
   \over \longrightarrow}\;  \bigoplus_{l=1}^{\gamma} \cO_\cM(M_l)\rightarrow 0\,,
\end{equation}
where the individual line bundles are restricted to the
complete intersection $\smash{\cM=\bigcap_{j=1}^c G_j}$. 
Here $r_{\mathcal V}$ additional fermionic gauge symmetries have been
introduced,
which for the Fermi superfields imply a deviation from chirality
$\overline{\cal D} \Lambda^a=\sqrt{2} \Sigma^i\, E_i{}^a$.
The additional neutral chiral superfields $\Sigma^i$ give rise
to an extra contribution to the scalar potential, which does not play any
role for our analysis. 
For more details on these  
fermionic gauge symmetries we refer  to \cite{Distler:1995mi,DistlerGreeneResolving}.
In the subsequent sections the notation 
\begin{equation}\label{eq_model configuration 2}
V_{N_1,\ldots, N_\delta}[M_1,M_2,\dots ,M_\gamma]\longrightarrow
\IP_{Q_1,\ldots ,Q_d}[S_1,\ldots ,S_c] 
\end{equation}
will be used for such a singular or smooth configuration. 
The constraints \eqref{eq_anomcancel} guarantee that the complete intersection
defines a threefold with vanishing first Chern class, i.e.~a Calabi-Yau manifold ${\cal M}$. In addition the vector bundle ${\cal V}$ is implied
to have $SU(n)$ structure group (if it is stable) and 
the $r(r-1)/2$ quadratic constraints imply the integrated Bianchi-identify
$c_2(V)=c_2(T)$ in each geometric phase.

The second phase arise for  $\xi < 0$. In this case $\langle p\rangle\ne 0$
with all other bosonic fields vanishing. 
Then,  the low-energy physics is 
described by a Landau-Ginzburg orbifold with a superpotential
\begin{equation}
 \mathcal W(X_i,\Lambda^a,\Gamma^j)= \sum_j \Gamma^j G_j(X_i) 
                      + \sum_a \Lambda^a F_a(X_i). 
\end{equation}
Methods have been developed to deal with  such  $(0,2)$ LG-models\cite{Distler:1993mk,Distler:1994hq},
which means in particular the generalization of 
the BRST methods for the computation
of the massless spectrum from $(2,2)$ LG orbifolds to the $(0,2)$ case.

It was first observed in \cite{DistlerKachruDuality}  
that in this superpotential the constraints 
$G_j$ and $F_a$ appear on equal footing, so that in particular
an exchange of them does not change the Landau-Ginzburg model as long as
all anomaly cancellation conditions are satisfied. In \cite{DistlerGreeneFeatures} this duality was
further investigated showing that this exchange is still possible after
resolving the generically singular base manifold. It is precisely
this duality we want to study in  this paper (see \cite{Adams:2003zy} for
another kind of $(0,2)$ duality).

%
\subsection{Geometric phases and GUT realizations}\label{sec_particle content}

Certain types of GLSMs are especially well suited for describing
the internal conformal field theory of four-dimensional
compactifications of the $E_8\times E_8$ heterotic string. 
Let us briefly discuss this for geometric phases
of GLSMs. 

Here the bosonic degrees of freedom take values in the
Calabi-Yau manifold ${\cal M}$ and their fermionic superpartners
couple to the pull-back of the rank three tangent bundle $T_{\cal M}$.
The left-moving fermions couple to the pull-back of
the vector bundle ${\cal V}$ with structure group $SU(n)$. 
The gauge group $G$ in the effective four-dimensional theory
is given by the commutant of $H=SU(n)$ in $E_8\times E_8$.
Embedding this into one of the two $E_8$ factors and
considering the other $E_8$ factor as  a hidden gauge symmetry,
one can directly get the canonical GUT gauge groups
\eq{
\label{gutgroups}
  &SU(3)\subset E_8 \Rightarrow G=E_6, \qquad SU(4)\subset E_8 \Rightarrow  G=SO(10)\\
   &SU(5) \subset E_8 \Rightarrow  G=SU(5)\; .
}
The massless matter particle content can then be determined
by computing corresponding vector bundle valued cohomology classes
on the Calabi-Yau threefold.
The respective classes can be read off from the decomposition
of $E_8$ into representations of $H\times G$.
For the three GUT cases \eqref{gutgroups} this is shown in table
\ref{table_GUT group representations via cohomology}.
\begin{table}[ht]
  \small
	\newcommand{\Xoplus}{\mllap{\oplus\,}}
  \begin{tabular}{l|cccccc}
    \# zero modes & &&&&&\\
    in reps of $H\times G$&  1 &   $h^1_{\cal M}({\cal V})$ & $h^1_{\cal M}({\cal V}^\ast)$ &  $h^1_{\cal M}(\Lambda^2 {\cal V})$ &  $h^1_{\cal M}(\Lambda^2 {\cal V}^\ast)$ & $h^1_{\cal M}({\cal V}\otimes {\cal V}^\ast)$  \\
    \hline\hline
    \qquad\;\;\,\parbox{1cm}{$E_8$\uspc \\ ${}\,\downarrow$} & \multicolumn{6}{|c}{\parbox{1cm}{248\uspc \\ ${}\;\downarrow$}}  \\ 
    $SU(3) \times E_6$     &  $(1,78)$ & $\Xoplus(3,27)$ & $\Xoplus(\overline{3},\overline{27})$  & & & $\Xoplus(8,1)$\uspc \\
    $SU(4) \times SO(10)$  & $(1,45)$ & $\Xoplus(4,16)$ & $\Xoplus(\overline{4},\overline{16})$ & $\Xoplus(6,10)$ & & $\Xoplus(15,1)$\\
    $SU(5) \times SU(5)$   &  $(1,24)$ & $\Xoplus(5,\overline{10})$ & $\Xoplus(\overline{5}, 10)$ & $\Xoplus(10,5)$ & $\Xoplus(\overline{10},\overline{5})$ & $\Xoplus (24,1)$\\
  \end{tabular}
  \caption{\small Matter zero modes in representations of the GUT group }
  \label{table_GUT group representations via cohomology}
\end{table}

\noindent
In this paper we are mainly concerned with $SU(3)$ bundles and therefore
observable gauge group $E_6$. In this case we get
chiral matter in the representations ${\bf 27}$ and ${\bf \overline{27}}$,
which are counted by $h^1_{\cal M}({\cal V})$ and
$h^1_{\cal M}({\cal V}^\ast)$, where by Serre duality the latter is equal to
$h^2_{\cal M}({\cal V})$. Moreover, let us mention that a necessary condition
for $\mu$-stability of the vector bundle ${\cal V}$ is
$h^0_{\cal M}({\cal V})=h^3_{\cal M}({\cal V})=0$.
In addition, the low-energy theory has massless gauge singlets,
which are counted by $h^1_{\cal M}({\cal V}\otimes {\cal V}^\ast)=h^1_{\cal
  M}({\rm End}({\cal V}))$. There are additional singlets related
to the complex structure and K\"ahler deformations of the Calabi-Yau
threefold, which are counted by $h^{2,1}({\cal M})$ and
$h^{1,1}({\cal M})$. Thus, one gets the total
number of
\eq{
\label{numberinfi}
{\rm D}({\cal M},{\cal V})= h^{1,1}({\cal M}) + h^{2,1}({\cal M}) +  h^{1}_{\cal M}({\rm
       End}({\cal V}))
}
massless gauge singlets. 
To determine the appearing  vector bundle valued cohomology classes
we employ the \cohomCalgKoszul\ implementation. As has been 
explained  in very much detail in \cite{Blumenhagen:2010ed}, this package is tailor
made for performing such computations for monad bundles over
complete intersections in toric varieties. Here we do not intend
to repeat the entire discussion, but just want to highlight a couple
of main issues:
\begin{itemize}
\item{The vector bundle is defined via a monad involving sums of line bundles. 
This monad can be split
into short exact sequences of vector bundles, which imply
long exact sequences in cohomology.}
\item{As input for the latter, one has to determine the cohomology
classes of line bundles ${\cal O}(D)$ over the Calabi-Yau manifold 
${\cal M}$, which is defined by the complete intersection
of hypersurfaces in the ambient toric variety $X$.
The hypersurface constraints can be considered as effective divisors 
$S_j\subset X$. Then, one has the so-called Koszul sequence
\beq\label{eq_koszulsequence}
  0 \fto \cO_X(D-S_J) \injto \cO_X(D) \surjto \cO_{S_j}(D) \fto 0,
\eeq
relating the line bundle on the hypersurface $S_j$ to line bundles
on  the ambient space $X$. This procedure can be iterated 
to eventually relate the line bundle on ${\cal M}$ to line
bundles on $X$.}
\item{Again short exact Koszul sequences imply long exact sequences
in cohomology. Thus, as final input data, one needs the cohomology
classes $H^i_X({\cal O}(D))$ of line bundles over the toric ambient
space. For this purpose, in \cohomCalgKoszul\ a fast algorithm 
was implemented which
was proposed in \cite{Blumenhagen:2010pv} and mathematically proven in 
\cite{2010arXiv1006.0780J,Rahn:2010fm}.}
\end{itemize} 

\noindent
Running through the exact sequences, generically one encounters
the problem that, in order to determine the dimension of 
certain cohomology classes, one has to determine the rank of
certain maps explicitly. Since this is a tedious and often
quite cumbersome exercise, in this paper we essentially
discard all cases where this happens and just stick to the ones,
where one has a sufficient number of zeros to determine
the dimensions of the appearing  cohomology classes uniquely.
It turns out that the latter cut can be made by using the following two  
assumptions, which imply   additional  zeros into the exact sequences:

\begin{itemize}
 \item We assume 
   stability of the  vector bundle $\mathcal V$. For generic
   monad bundles this is in
   general difficult  to check. It implies 
\eq{
\label{esiwienb}
        h^0_\cM(\mathcal V) = 0\, , &\qquad  h^3_\cM(\mathcal V)=0\, , \\
       h^0_\cM(\mathcal V \otimes \mathcal V^\ast) = 1\, , &\qquad  h^3_\cM(V
       \otimes \mathcal V^\ast)= 1\; .
}
 \item The computation of  $h^i_{\cal M}({\rm End}({\cal V}))$ involves  a map, for which we assume that it is surjective. 
This map $\varphi$  appears as the second map in the exact sequence
\eq{
\label{esiwien}
  0 \longrightarrow H^0\Bigl({\cal E}_{\cal M}^\ast \otimes \bigoplus_{l=1}^{\gamma}
  \cO_{\cM}(M_l)\Bigr)&\longrightarrow
  H^0\Bigl(\bigoplus_{a=1}^{\delta}\bigoplus_{l=1}^{\gamma}
  \cO_{\cM}(M_l-N_a)\Bigr)\\
  & \stackrel{\varphi}{\surjto} H^0\Bigl(\bigoplus_{j=1}^{c} \cO_{\cM}(M_j)^{\oplus r_{\mathcal V}}\Bigr) \longrightarrow \cdots
}     
which arises as an intermediate step in the 
long exact sequences in cohomology, after  writing ${\cal V}\otimes {\cal
  V}^\ast$ via short exact sequences (see \cite{Distler:1988jj,
  RalphTargetSpace1}  for more details). 
We  actually checked for quite a few examples that this holds, but
do not have a proof that  generally this is the case.
\end{itemize}

\vspace{0.2cm}
\noindent
Finally, we comment on the $(0,2)$ moduli space.
The number of first order deformations \eqref{numberinfi}
is not necessarily equal to the true dimension
${\rm dim}({\cal D}({\cal M},{\cal V}))$ of the total moduli space of the 
theory, as there can be obstructions.
Mathematically, this means that there can be complex structure
deformations, under which the bundle cannot be kept holomorphic\footnote{
As explained in the physical context for instance in \cite{Anderson:2010mh},
this is captured by the so-called Atiyah-class.}.
Physically, this is described by the tree-level four-dimensional 
superpotential
\eq{
           W=\int_{\cal M}  \Omega_3\wedge \omega_{\rm YM}
}
where $\Omega_3$ denotes the holomorphic $(3,0)$ form on the Calabi-Yau
and $\omega_{\rm YM}={\rm tr}(A\wedge  dA -{\frac{2i}{3}} A \wedge A\wedge A)$ 
the Chern-Simons 
form of the $SU(n)$ gauge connection $A$. The flat directions of the scalar potential
induced by $W$ define the true moduli space of the configuration
$({\cal M},{\cal V})$.  
A non-renormalization theorem states that,
beyond this leading order contribution, there can only be non-perturbative
corrections from world-sheet instantons. For more information on this
important issue, we refer to the literature \cite{Distler:1987ee,Silverstein:1995re,Basu:2003bq,Beasley:2003fx,Aspinwall:2010ve,Aspinwall:2011us}.

Unfortunately, the superpotential is 
hard to compute for a concrete $(0,2)$ model $({\cal M},{\cal V})$. 
However, we know that at least the independent complex coefficients
in the holomorphic sections $G_j$ and $F_a{}^l$, i.e. the toric
deformations, keep the vector bundle holomorphic. 

\section[Explicit construction of dual (0,2) models]{Explicit construction of dual \boldmath{$(0,2)$} models}\label{sec_construction}
%
In this section we further generalize the analysis
of $(0,2)$ target space dualities presented in  
\cite{DistlerKachruDuality,DistlerGreeneFeatures,RalphTargetSpace1,RalphTargetSpace2}
and propose a general procedure that can be used to generate dual models from
almost any monad over a complete intersection Calabi-Yau base space, not necessarily
endowed with a Landau-Ginzburg phase. In particular,  one can show that performing this procedure, the anomaly cancellation conditions remain satisfied for the dual models.

\subsection{Outline of the generic construction of dual models}\label{subsec_Idea of the generic construction of dual models}

We will use the following notation for the $U(1)$ charges and (multi)degrees of fields and homogeneous functions, respectively
$$
\text{charge}(\text{field }X) =: ||X|| \quad \text{and} \quad \text{degree}(\text{function }G) =: ||G||\,.
$$
Before we show how to construct dual $(0,2)$ models explicitly, let us
outline the generic procedure.
We will start with a smooth $(0,2)$ model $({\cal M},{\cal V})$ for which 
all anomaly cancellation conditions \eqref{eq_anomcancel} are satisfied.
Using the existence of non-geometric phases, where some of the bosonic fields
$p_l$   receive a vev, we perform an exchange  of some of the Fermi superfields
and the corresponding polynomials. The resulting new GLSM is claimed
to be target space dual to the initial one. 
A necessary condition is that the massless charged matter spectrum
and the generic number of massless gauge singlets  ${\rm D}({\cal M},{\cal V})$ 
should be identical.
Determining  these numbers  in the geometric phases
of a dual pair, they should agree. More concretely, we follow
the procedure:

\paragraph{The procedure:}
\begin{enumerate}
 \item Construct the GLSM phases of a smooth $(0,2)$ model
    $({\cal M},{\cal V})$.
 \item Go to a phase where one of the $p_{l}$, say $p_{1}$, is not allowed to vanish and hence obtains a vev $\left< p_{1} \right>$.
 \item Perform a rescaling of $k$ Fermi superfields by the constant 
vev $\left< p_{1} \right>$ and exchange the role 
of some $\Lambda^a$ and $\Gamma^j$
       $$\tilde\Lambda^{a_i} :={\frac{\Gamma^{j_i}}{\left< p_{1} \right>}}, \quad
\tilde\Gamma^{j_i}  := \left< p_{1} \right>\Lambda^{a_i}\,,\quad ~\forall i=1,...,k\,,$$
      with  $\sum_{i} || G_{j_i} || = \sum_{i} || F_{a_i}{}^{1} ||$ for anomaly cancellation.
 \item Move to a region in the bundle moduli space where the $\Lambda^{a_i}$
   only appear in terms with $P_{1}$ for all $i$. This means that  
       we choose the coefficients in the bundle defining polynomials 
$F_{a}{}^l$ such that $$F_{a_i}{}^l=0\, , 
  \qquad\forall~l\neq 1,\ ~i=1,...,k\,.$$
 \item Leave the non-geometric phase and define the
   Fermi superfields of the new GLSM 
such that each term in the superpotential is $U(1)^r$ gauge invariant. This means
 $$||\tilde\Lambda^{a_i}|| = ||\Gamma^{j_i}||-||P_{1}|| \quad\text{and}\quad
||\tilde\Gamma^{j_i}|| = ||\Lambda^{a_i}|| + ||P_{1}||\, . $$
 \item Returning to a generic point in moduli space defines a new dual 
$(0,2)$ GLSM which in a geometric phase corresponds to a different
 Calabi-Yau/vector bundle configuration $(\widetilde {\cal M},\widetilde {\cal V})$.
\end{enumerate}
%
\subsection{Explicit procedure}\label{subsec_construction of dual models}
Let us now be  more explicit and show  how this procedure works in detail. 
For presentational purpose, we will restrict ourselves to the choice $k=2$.
This is also the case used in  performing  the $(0,2)$ 
landscape analysis   to be reported on 
in section \ref{sec_landscape studies}.

Let us consider a holomorphic vector bundle ${\cal V}$, 
obtained from a monad over a base Calabi-Yau manifold ${\cal M}$,
which we denoted as 
\begin{equation}
V_{N_1,...,N_\delta}[M_1,...,M_\gamma]\longrightarrow \IP_{Q_1,...,Q_d}[S_1,...,S_c]\,.
\end{equation}
It is also assumed that the anomaly cancellation conditions
\eqref{eq_anomcancel} are satisfied.
We now require that for one specific $M_{l_0}$ 
there exist two $N_{a_j}$'s such that $N_{a_j}^{(\alpha)} < M_{l_0}^{(\alpha)}$.
Let us  choose, without loss of generality, $l_0=1$ and rearrange the 
$N's$ in the monad such that they are the first $2$. Thus, we have
\begin{equation}
V_{N_1,...,N_\delta}[M_1,...,M_\gamma]\longrightarrow \IP_{Q_1,...,Q_d}[S_1,...,S_c]\, 
\end{equation}
and the corresponding superpotential has the form
\begin{eqnarray}\label{eq_superpotential expanded}
\mathcal W =\sum_{j=1}^{c}\Gamma^j G_j + \sum_{a=1}^{2} P_1 \Lambda^a F_a{}^1 +  \sum_{a=3}^{\delta} P_1 \Lambda^a F_a{}^1 + \sum_{l=2}^{\gamma} \sum_{a=1}^{\delta} P_l\, \Lambda^a F_a{}^l\,.
\end{eqnarray}
For the case $\gamma=1$, i.e.~a monad $V_{N_1,...,N_\delta}[M_1]$, 
the last term would be absent and
the GLSM features  a Landau-Ginzburg phase in which  $p_1$
carries  a vacuum expectation value. For the case $\gamma=2$, i.e.~$V_{N_1,...,N_\delta}[M_1,M_2]$,  with
only a single  $U(1)$ gauge symmetry,  even though there is no Landau-Ginzburg
phase anymore, one may still find a phase in which $p_1$ and $p_2$ cannot vanish
simultaneously. This describes a Landau-Ginzburg model fibered over a
$\IP^1$,  parametrized by the homogeneous coordinates $(p_1,p_2)$. 
Thus, $p_1$ or $p_2$ are not allowed  to vanish simultaneously.

Hence, for these two simple cases, 
one can explicitly identify a phase, in which  
not all vevs $\langle p_l\rangle$ do vanish.
Our dual model generating algorithm starts on a sublocus 
where a specific vev $\langle p_1\rangle\ne 0$. This is all we need
to perform the desired change of variables.
However,  since this is a tedious analysis, for the automated landscape study 
in section \ref{sec_landscape studies}, we did not check the existence
of such a phase for each
individual case, but proceeded under the assumption that
it  exists.

Considering \eqref{eq_superpotential expanded} and comparing the first sum
with the second one, one realizes  that they only differ by the additional
chiral superfield $P_1$. If one now  goes into the aforementioned phase, 
where $p_1$ obtains a vev, the effective superpotential becomes 
\begin{eqnarray}\label{eq_superpotential expanded1}
\mathcal W = \sum_{j=1}^{c}\Gamma^j G_j + \sum_{a=1}^{2} \left< p_1 \right> \Lambda^a F_a{}^1 +  \sum_{a=3}^{\delta} \left< p_1 \right> \Lambda^a F_a{}^1 + \sum_{l=2}^{\gamma} \sum_{a=1}^{\delta} P_l\, \Lambda^a F_a{}^l\,.
\end{eqnarray}
Now, we want to perform an exchange  of two of the Fermi superfields appearing
in the first and the second term  \eqref{eq_superpotential expanded1}. 
Without loss of generality we choose these two pairs  to be  
$\Gamma^1,~\Gamma^2$ and $\Lambda^1,~\Lambda^2$. 
For this purpose, we first need to  move to a region in the bundle moduli 
space, where  the sections $F_a{}^l$ satisfy 
\eq{
F_1{}^l = F_2{}^l = 0\,  \qquad \forall~l\neq 1\,.
}
This guarantees that the superpotential takes the restricted form
\begin{eqnarray}\label{eq_superpotential expanded2}
\mathcal W\!\! &=&\!\! \sum_{j=1}^{c}\Gamma^j G_j + \left< p_1 \right>
\Lambda^1 F_1{}^1 + \left< p_1 \right> \Lambda^2 F_2{}^1
+ \parbox{4cm}{\center terms independent\\ of
  $\Gamma^j,~\Lambda^1,\Lambda^2$}\!\! .
\end{eqnarray}
Now the superfields  $\Gamma^j$ and  $\Lambda^1,~\Lambda^2$ appear on 
an equal footing and hence do the homogeneous functions $G_j$, $F_1{}^1$ and
$F_2{}^1$. Thus, in this non-geometric phase, their 
distinctive geometric origin  as   hypersurface
constraints $G_j$ and sections $F_1{}^1$ defining
the bundle is completely lost. 

Not every such exchange of  $\Gamma^1,~\Gamma^2$ and $\Lambda^1,~\Lambda^2$ 
leads to a fully fledged new $(0,2)$ GLSM, after moving
away from this special point in moduli space.
For a GLSM  the anomaly cancellation conditions \eqref{eq_anomcancel} 
have to be satisfied. In the following we will describe 
two different scenarios. 
The first one corresponds to a consistent exchange of $F$'s and $G$'s where 
\eq{
 ||G_1||+||p_1|| \neq 0 \quad \text{and} \quad ||G_2||+||p_1|| \neq 0\,,
}
while in the second scenario we will have the situation where
\eq{
 ||G_1||+||p_1|| = 0 \quad \text{and} \quad ||G_2||+||p_1|| \neq 0\,.
}
The latter naively  leads to Fermi superfields
of vanishing charge. We will see that this is not really the case, but 
that instead for the dual model the number of $U(1)$ gauge symmetries
gets enlarged. Thus, in the geometric phase the dimension
of the K\"ahler moduli space increases.
We will find that the Fermi superfield is actually  charged under this 
additional $U(1)$ gauge group.

\paragraph{Dual models with equal number of \boldmath{$U(1)$} actions:}

If we want to consistently exchange $F$'s and $G$'s, we have to make sure that
the linear  anomaly cancellation condition remains satisfied. For the exchange of two of them, say 
\eq{
F_1{}^1,~F_2{}^1 \quad \leftrightsquigarrow \quad G_1,~G_2\,,
}
this requires the following relation of their homogeneous multi-degrees:
\begin{equation}\label{eq_F and G condition}
||F_1{}^1|| + ||F_2{}^1|| = ||G_1|| + ||G_2|| \quad \Rightarrow \quad 2M_1 - N_1 - N_2 = S_1 + S_2\,.
\end{equation}
As long as $||G_1||,~||G_2||$ both are  not equal to $M_1$, we can perform this exchange without any problem. If there is a phase where $p_1$ is not allowed to vanish, we can write the effective superpotential at low energies by integrating out $p_1$ and moving to the corresponding region in moduli space as seen in \eqref{eq_superpotential expanded2}. To make the exchange of the homogeneous polynomials manifest, we have to absorb this vev by a rescaling of some of the fields. We obtain the new configuration as
\begin{eqnarray}\label{eq_dual superpotential} 
\begin{aligned}
\mathcal W = &~\tilde\Gamma^1 \tilde G_1 + \tilde\Gamma^2 \tilde G_2 + \sum_{j=3}^{c}\Gamma^j G_j + \\
  &~\left<p_1\right> \tilde\Lambda^1 \tilde F_1{}^1 + \left<p_1\right> \tilde\Lambda^2 \tilde F_2{}^1 + \sum_{a=3}^{\delta} \left<p_1\right> \Lambda^a F_a{}^1 +
    \sum_{l=2}^{\gamma} \sum_{a=3}^{\delta} P_l\, \Lambda^a F_a{}^l\,,
\end{aligned}
\end{eqnarray}
where we performed rescalings
\begin{eqnarray}\label{eq_dual quantities 1}
\begin{aligned}
 \tilde\Gamma^1 &:=& \left<p_1\right> \Lambda^1,\quad \tilde\Gamma^2 &:=& \left<p_1\right> \Lambda^2,\quad
    \tilde \Lambda^1 &:=& \frac{\Gamma^1}{\left<p_1\right>},\quad \tilde \Lambda^2 &:=& \frac{\Gamma^2}{\left<p_1\right>},\\
 \tilde G_1 &:=& F_1{}^1,\quad \tilde G_2 &:=& F_2{}^1,\quad \tilde F_1{}^1 &:=& G_1,\quad \tilde F_2{}^1 &:=& G_2 \,.
\end{aligned}
\end{eqnarray}
This superpotential \eqref{eq_dual superpotential} is identical to the initial
one, but arises from a completely different GLSM. At this point we can see that
it was essential to move to a specific region of the moduli space, as  the
rescaling \eqref{eq_dual quantities 1} would not have been consistent, if there
were terms like $\Lambda^1 F_1{}^2$. Since the homogeneous polynomial
$F_1{}^2$ might not have the same multi-degree as $F_1{}^1$, 
the rescaling \eqref{eq_dual quantities 1} would give
rise to a term in the superpotential which is not gauge invariant.

The new charges and degrees of the superfields in the GLSM 
read
\begin{eqnarray}
\label{eq_dual configuration 1}
V_{\tilde{N}_1,\tilde{N}_2,N_3,...,N_\delta}[M_1,M_2,...,M_\gamma]
\longrightarrow
\IP_{Q_1,...,Q_d}[\tilde{S}_1,\tilde{S}_2,S_3,...,S_c]\,,
\end{eqnarray}
with
\begin{eqnarray}
 \quad \tilde{N}_1:=M_1-S_1,\quad \tilde{N}_2:=M_2-S_2,\quad\tilde{S}_1:=||F_1{}^1||,\quad\tilde{S}_2:=||F_2{}^1||\,.
\end{eqnarray}
We prove in appendix \ref{app_anomaly cancellation} that this $(0,2)$ GLSM
fulfills all anomaly cancellation conditions and hence defines a
genuine new model.
In particular, for the new model one can consider
generic points in the moduli space and perform its own phase analysis,
i.e.~consider the total complexified K\"ahler moduli space.
This also includes  the large volume limits of potential  geometric phases. 
There, it describes now  topologically
distinct Calabi-Yau  manifolds equipped with  different vector bundles over 
them.

We were calculating  various  examples of this kind and
found that the following intriguing relation holds
in over  90\% of them \footnote{In going
through the various Koszul sequences arising for 
determining the bundle deformations, we were assuming
the surjectivity of the map $\varphi$ in  \eqref{esiwien}. 
Moreover, we were also blindly  assuming that the new bundle 
$\widetilde {\cal V}$ is 
$\mu$-stable over the new base manifold $\tilde {\cal M}$ \eqref{esiwienb}.
We expect that a  mismatch  merely indicates that for this
specific example one of these assumptions is violated.}:
\begin{eqnarray}
\begin{aligned}
\label{eq_match chiral spectrum}
h_{{\cal M}}^\bullet({\cal V}) &= h_{\widetilde{\cal  M}}^\bullet(\widetilde
{\cal V}) \\
\label{eq_match moduli space}
 h_{{\cal M}}^{1,1}+h_{{\cal M}}^{2,1}+h_{{\cal M}}^1 (\text{End}({\cal V}))
 &= h_{\widetilde {\cal M}}^{1,1}+h_{\widetilde {\cal M}}^{2,1}+h_{\widetilde
   {\cal M}}^1 (\text{End}(\widetilde {\cal V}) )\,.
\end{aligned}
\end{eqnarray}
This means  that, at least on a dimensional basis, the chiral spectra as well
as the number of massless singlets  of the two $(0,2)$ models, 
$({\cal M},{\cal V})$ and $(\widetilde{\cal M},\widetilde {\cal V})$, agree.
From the rescalings  \eqref{eq_dual quantities 1}, 
it is clear that the moduli space of $({\cal M},{\cal V})$ is related
to the moduli space of $(\widetilde{\cal
  M},\widetilde {\cal V})$ by  an exchange of
complex structure  and bundle moduli. The K\"ahler
moduli space was rather untouched. 

In the following we will describe a way to
construct dual $(0,2)$ models in which the Picard group 
of the ambient space becomes
larger so that   also the K\"ahler moduli spaces are non-trivially involved
in the duality.

\paragraph{Dual models with an additional \boldmath{$U(1)$} action.}
Let us start again with the monad of the model  
\eqref{eq_model configuration 2} and pick two specific maps $F_a{}^l$, e.g.
$F_1{}^1$  and $F_2{}^1$, belonging to
$M_1$. Now choose one of the hypersurfaces defining our base, say $S_1$ such
that we can find positive degrees $B^{(\alpha)}$ for all $\alpha$  satisfying
the multi-degree 
equation
\begin{equation}\label{eq_definition of B}
B = ||F_1{}^1|| + ||F_2{}^1|| - S_1\,.
\end{equation}
We can now introduce a new coordinate $y_1$ with multi-degree $B$ and also a
new hypersurface $G^B$ described by a homogeneous polynomial of multi-degree
$B$. This means we simply introduce a new Fermi superfield along with a new
chiral superfield that have opposite charges. Doing that at the same time
does not cause any changes to our model $({\cal M},{\cal V})$ 
and we can express it  as
\begin{equation}
V_{N_1,...,N_\delta}[M_1,...,M_\gamma]\longrightarrow \IP_{Q_1,...,Q_d,B}[S_1,...,S_c,B]\,.
\end{equation}
In case that $S_1\neq M_1$ and $B\neq M_1$ we can proceed in the same way as
in the paragraph above and redefine fields as described there. 
Thus,  we arrive at the following configuration:
\begin{eqnarray}
V_{\tilde{N}_1,\tilde{N}_2,N_3,...,N_\delta}[M_1,M_2,...,M_\gamma]
\longrightarrow
\IP_{Q_1,...,Q_d,B}[\tilde{S}_1,S_2,...,S_c,\tilde B]\,,
\end{eqnarray}
where
\begin{eqnarray}
 \quad \tilde{N}_1:=M_1-S_1,\quad \tilde{N}_2:=M_2-B,\quad\tilde{S}_1:=||F_1{}^1||,\quad\tilde{B}:=||F_2{}^1||\,.
\end{eqnarray}
The only new
issue is that, as we have introduced a new coordinate $y_1$ with
multi-degree $B$, we might get   new singularities 
in the dual model $(\widetilde{\cal M},\widetilde {\cal V})$.
These need  to be resolved before performing
calculations in the large volume limit. In addition, after the resolution 
of the base, we also have to resolve the bundle  without spoiling the
anomaly cancellation conditions. How this can be done has
been explained in \cite{DistlerGreeneResolving}.

Applying this procedure for instance to the tangent bundle, i.e.~${\cal V}=T_{\cal M}$, we encounter the situation  that
\eq{
S_1 = M_1\quad (\text{or equivalently } B=M_1) \; .
} 
Now doing the same steps as in the last paragraph,
 we arrive at the following configuration:
\begin{eqnarray}
\label{eq_dual configuration 2}
V_{\tilde{N}_1,\tilde{N}_2,N_3,...,N_\delta}[M_1,M_2,...,M_\gamma]
\longrightarrow
\IP_{Q_1,...,Q_d,B}[\tilde{S}_1,S_2,...,S_c,\tilde B]\,,
\end{eqnarray}
where
\begin{eqnarray}\label{eq_dual configuration 2.2}
 \quad \tilde{N}_1:=M_1-S_1 = \vec{0},\quad \tilde{N}_2:=M_2-B,\quad\tilde{S}_1:=||F_1{}^1||,\quad\tilde{B}:=||F_2{}^1||\,.
\end{eqnarray}
Thus, in the new model we find the Fermi superfield  
$\tilde \Lambda^1 := \frac{\Gamma^1}{\left<p_1\right>}$ to be uncharged 
under all of our $U(1)$ symmetries\footnote{
It was argued in \cite{DistlerGreeneFeatures} that
one can employ one of the $r_{\mathcal V}$ additional 
fermionic gauge symmetries given in
\eqref{eq_general monad} in order to gauge it away. This works fine for their
example but it cannot be used 
for our case, where the newly introduced field is them self uncharged under all $U(1)$'s.}.

To proceed, we introduce an additional $U(1)$ gauge symmetry
under which the former uncharged Fermi superfield  
carries a non-vanishing charge.
We do that by a formal blow-up of a $\IP^1$ with coordinates $y_1,y_2$ 
so that the charges of the resulting GLSM read
\begin{table}[ht]
\begin{center}
\begin{tabular}{c}
%
\begin{tabular}{| c   c   c   c   c || c   c   c  c| }
  \hline                       
   $x_1$ & ... & $x_d$ & $y_1$ & $y_2$  &$\Gamma^1$ & ... & $\Gamma^c$ & $\Gamma^B$ \\
 \noalign{\hrule height 1pt}
     0   & \dots &   0   &   1   &    1       &    $0$     & ... &     0      &    $-1$      \\
   $Q_1$ & \dots & $Q_d$ &   $B$ &    0       & $-S_1$   & ... & $-S_c$     &     $-B$     \\
  \hline  
\end{tabular}
\hfill
\vspace{0.3cm}
\\
%
\begin{tabular}{| c  c  c  c    || c  c  c  c |}
  \hline                       
   $\Lambda^1$ & $\Lambda^1$ & \dots & $\Lambda^{\delta}$ & $p_1$ & $p_2$ &... & $p_\gamma$ \\\noalign{\hrule height 1pt}
   0   & 0   & \dots & 0 & $-1$ & 0 & \ldots & 0  \\
   $N_1$ & $N_{2}$ & \ldots & $N_{\delta}$ & $-M_1$ & $-M_2$ &  ... & $-M_\gamma$  \\
  \hline  
\end{tabular}
\hfill
\end{tabular}
\end{center}
\label{tab_modified degrees of all fields}
\end{table}

\noindent
This configuration is equivalent to  the initial 
one. Just eliminate the coordinate $y_1$ via the constraint
$G^B=y_1=0$ and use the additional $U(1)$ gauge symmetry and 
the corresponding D-term constraint
to fix $y_2$ to a real constant. Then the geometry 
reduces to $\IP_{Q_1,...,Q_d}[{S}_1,S_2,...,S_c]\times {\rm pt}$.
So the new configuration also satisfies the anomaly cancellation conditions.

Applying the exchange of $G$'s and $F$' as in \eqref{eq_dual configuration 2},
\eqref{eq_dual configuration 2.2},  the former uncharged Fermi superfield  
in the dual configuration now carries a non-zero charge under the new $U(1)$.
This is what we wanted to achieve and in particular allows us
to systematically generate dual models of the heterotic 
$({\cal M},T_{\cal M})$ models.
The data of the GLSM of the dual configuration are listed below, for which
it is proven in appendix \ref{app_anomaly cancellation} that they  still 
satisfy all anomaly cancellation conditions \eqref{eq_anomcancel}:
\begin{table}[h]
\begin{center}
\begin{tabular}{c}
\begin{tabular}{| c  c  c  c  c || c  c  c  c  c | }
  \hline                       
   $x_1$ & \ldots & $x_d$ & \boldmath{$y_1$} & \boldmath{$y_2$} &\boldmath{$\tilde\Gamma^1$} & $\Gamma^2$ & ... & $\Gamma^c$ & \boldmath{$\tilde\Gamma^B $}\\
 \noalign{\hrule height 1pt}
   $ 0 $ & \dots & $ 0 $ & \boldmath{$1$}  & \boldmath{$1$} & \boldmath{$-1$}
  & $ 0  $ & \ldots &   0    & \boldmath{$-1$}\\
   $Q_1$ & \ldots & $Q_d$ & \boldmath{$B$} & \boldmath{$0$} & \boldmath{$-(M_1-N_1)$} & $-S_2$ & \ldots & $-S_c$ & \boldmath{$-(M_1-N_2)$} \\
  \hline  
\end{tabular}
\hfill
\vspace{0.3cm}
\\
%
\begin{tabular}{| c  c  c  c || c  c  c  c |}
  \hline
   \boldmath{$\tilde{\Lambda}_1$} & \boldmath{$\tilde\Lambda^2$} & \ldots & $\Lambda^{\delta}$ & \boldmath{$p_1$} & $p_2$ & \ldots & $p_\gamma$ \\
 \noalign{\hrule height 1pt}
   \boldmath{$1$} & \boldmath{$0$} & \ldots &   $0$     & \boldmath{$-1$} & $ 0 $  & \dots &
 $ 0 $  \\
   \boldmath{$0$} & \boldmath{$M_2 -B$} & \ldots & $N_{\delta}$ & \boldmath{$-M_1$} & $-M_2$ & \ldots & $-M_\gamma$  \\
  \hline  
\end{tabular}
\hfill\hfill
\end{tabular}
\end{center}
\label{table_new degrees of all fields}
\end{table}

\vspace{0.5cm}

\subsection{Example}\label{subsec_example1}

Let us present  a comparably  simple example.
Consider the $(0,2)$ model
\eq{
V_{1,1,1,1,2,2,2}[3,4,3]\longrightarrow \IP_{1,1,1,1,2,2,2}[3,4,3] \,.
}
where the vector bundle is simply a deformation of the tangent 
bundle of the Calabi-Yau manifold ${\cal M}$.
Since this configuration is singular, we have to resolve it by introducing a
new coordinate. After some reordering of the bundle data, this yields 
the following smooth configuration:
%
%
\begin{equation}
\begin{aligned}
&\begin{array}{|c||c|}
\hline
x_i & \Gamma^j \\
\noalign{\hrule height 1pt}
\begin{array}{cccccccc}
 0 & 0 & 0 & 0 & 1 & 1 & 1 & 1 \\
 1 & 1 & 1 & 1 & 2 & 2 & 2 & 0
\end{array}
&
\begin{array}{ccc}
 -1 & -2 & -1 \\
 -3 & -4 & -3
\end{array}\\
\hline
\end{array}
\\[0.1cm]
&\begin{array}{|c||c|}
\hline
\Lambda^a & p_l \\
\noalign{\hrule height 1pt}
\hline
\begin{array}{cccccccc}
 0 & 1 & 1 & 1 & 0 & 0 & 0 & 1 \\
 1 & 2 & 2 & 2 & 1 & 1 & 1 & 0
\end{array}
&
\begin{array}{ccc}
 -1 & -2 & -1 \\
 -3 & -4 & -3
\end{array}\\
\hline
\end{array}\,.
\end{aligned}
\end{equation}

\normalsize
\noindent
We now compute  the number of chiral matter zero modes 
as well as the number of massless singlets  ${\rm D}({\cal M},{\cal V})$:
\eq{
\label{eq_match chiral spectrum example 1}
h_{\cal M}^\bullet({\cal }V) &= ( 0, 68 , 2,0)\,,\\
 h_{\cal M}^{1,1}+h_{\cal M}^{2,1}+h_{\cal M}^1 (\text{End}({\cal V})) &=  2 + 68 + 140 = 210\,.
}
Now we can use the procedure from last section. First we introduce a new field
$y_1$ which is not charged under the $U(1)$'s we have so far,
and introduce a new hypersurface which is also neutral. 
We  formally get the following set of data
\begin{equation}
\begin{aligned}
&\begin{array}{|c||c|}
\hline
x_i & \Gamma^j \\
\noalign{\hrule height 1pt}
\begin{array}{ccccccccc}
 0 & 0 & 0 & 0 & 1 & 1 & 1 & 1 & 0\\
 1 & 1 & 1 & 1 & 2 & 2 & 2 & 0 & 0
\end{array}
&
\begin{array}{cccc}
 -1 & -2 & -1 & -0 \\
 -3 & -4 & -3 & -0
\end{array}\\
\hline
\end{array}
\\[0.1cm]
&\begin{array}{|c||c|}
\hline
\Lambda^a & p_l \\
\noalign{\hrule height 1pt}
\begin{array}{cccccccc}
 0 & 1 & 1 & 1 & 0 & 0 & 0 & 1 \\
 1 & 2 & 2 & 2 & 1 & 1 & 1 & 0
\end{array}
&
\begin{array}{ccc}
 -1 & -2 & -1 \\
 -3 & -4 & -3
\end{array}\\
\hline
\end{array}\,.
\end{aligned}
\end{equation}
Notice that the homogeneous functions $F_3{}^2$ and $F_4{}^2$ 
are both of the same multi-degree, 
\eq{
||F_3{}^2||=||F_4{}^2|| =-||p_2|| - ||\Lambda^3||=-||p_2|| - ||\Lambda^4|| =
\left(\!\!\begin{array}{c}2-1 \\ 4-2\end{array}\!\!\right) = \left(\!\!\begin{array}{c}1
    \\ 2\end{array}\!\!\right)
}
and hence we can exchange the new hypersurface together with $G_2$ with 
these two, satisfying 
\eq{
 ||F_3{}^2|| + ||F_4{}^2|| =  \left(\!\begin{array}{c}1 \\ 2\end{array}\!\right) + \left(\!\begin{array}{c}1 \\ 2\end{array}\!\right) =  \left(\!\begin{array}{c}2 \\ 4\end{array}\!\right)+ \left(\!\begin{array}{c}0 \\ 0\end{array}\!\right) = S_2 + B \;.
}
From the last section,  we know how to exchange these functions and how 
to redefine the $\Lambda$'s and $\Gamma$'s in order to obtain 
a sensible new monad.  Namely we perform the rescalings
\begin{eqnarray}
\begin{aligned}
 \tilde\Gamma^2 &:=& \left<p_2\right> \Lambda^3,\quad \tilde\Gamma^B &:=& \left<p_2\right> \Lambda^4,\quad
    \tilde \Lambda^3 &:=& \frac{\Gamma^2}{\left<p_1\right>},\quad \tilde \Lambda^4 &:=& \frac{\Gamma^B}{\left<p_2\right>},\\
 \tilde G_2 &:=& F_3{}^2,\quad \tilde G_B &:=& F_4{}^2,\quad \tilde F_3{}^2 &:=& G_2,\quad \tilde F_4{}^2 &:=& G_B \,,
\end{aligned}
\end{eqnarray}
yielding the effective superpotential
\begin{eqnarray}\label{eq_dual superpotential example 1} 
\begin{aligned}
\mathcal W = &~\tilde\Gamma^2 \tilde G_2 + \tilde\Gamma^B \tilde G_B + \sum_{j=1,3}\Gamma^j G_j + \\
  &\qquad ~\left<p_2\right> \tilde\Lambda^3 \tilde F_3{}^2 + \left<p_2\right> \tilde\Lambda^4 \tilde F_4{}^2 +
    \sum_{l=1,3} \sum_{a\neq 3,4} P_l \Lambda^a F_a{}^l\,.
\end{aligned}
\end{eqnarray}
The new charges of the constructed model read
\begin{eqnarray}
\begin{aligned}
 ||\tilde\Gamma^2|| &=&\!\!\!\!   \left(\!\begin{array}{c}-1 \\ -2\end{array}\!\right) ,\quad ||\tilde\Gamma^B|| &=&\!\!\!\! \left(\!\begin{array}{c}-1 \\ -2\end{array}\!\right),\quad
    ||\tilde \Lambda^3|| &=&\!\!\!\! \left(\!\begin{array}{c}0 \\ 0\end{array}\!\right) ,\quad ||\tilde \Lambda^4|| &=&\!\!\!\! \left(\!\begin{array}{c}2 \\ 4\end{array}\!\right),\\
 ||\tilde G_2|| &=&\!\!\!\! \left(\!\begin{array}{c}1
    \\ 2\end{array}\!\right),\quad ||\tilde G_B|| &=&\!\!\!\! \left(\!\begin{array}{c}1 \\ 2\end{array}\!\right),\quad 
\tilde ||F_3{}^2|| &=&\!\!\!\! \left(\!\begin{array}{c}2 \\ 4\end{array}\!\right),\quad ||\tilde F_4{}^2|| &=&\!\!\!\! \left(\!\begin{array}{c}0 \\ 0\end{array}\!\right) \,.
\end{aligned}
\end{eqnarray}
We realize that one of the new $\Lambda$'s is uncharged under both
$U(1)$'s. Thus,  we introduce a new $U(1)$ along with a new coordinate 
in the base, which gives all new fields a charge. 
Doing that in the way explained in
the last section and going back to a generic point in moduli space, we arrive
at the GLSM 
\begin{equation}
\begin{aligned}
&\begin{array}{|c||c|}
\hline
x_i & \Gamma^j \\
\noalign{\hrule height 1pt}
\begin{array}{cccccccccc}
 0 & 0 & 0 & 0 & 0 & 0 & 0 & 0 & 1 & 1 \\
 0 & 0 & 0 & 0 & 1 & 1 & 1 & 1 & 0 & 0 \\
 1 & 1 & 1 & 1 & 2 & 2 & 2 & 0 & 0 & 0
\end{array}
&
\begin{array}{cccc}
 -0 & -1 & -0 & -1 \\
 -1 & -1 & -1 & -1 \\
 -3 & -2 & -3 & -2
\end{array}\\
\hline
\end{array}
\\[0.1cm]
&\begin{array}{|c||c|}
\hline
\Lambda^a & p_l \\
\noalign{\hrule height 1pt}
\begin{array}{cccccccc}
 0 & 0 & 1 & 0 & 0 & 0 & 0 & 0 \\
 0 & 1 & 0 & 2 & 0 & 0 & 0 & 1 \\
 1 & 2 & 0 & 4 & 1 & 1 & 1 & 0
\end{array}
&
\begin{array}{ccc}
 -0 & -1 & -0 \\
 -1 & -2 & -1 \\
 -3 & -4 & -3
\end{array}\\
\hline
\end{array}\,.
\end{aligned}
\end{equation}
As was generically shown, this configuration satisfies the conditions \eqref{eq_anomcancel} and we obtain the following topological data:
\eq{
\label{eq_match chiral spectrum 2}
h_{\widetilde{\cal M}}^\bullet(\widetilde{\cal V}) &= ( 0, 68 , 2, 0)\,,\\
 h_{\widetilde{\cal M}}^{1,1}+h_{\widetilde{\cal M}}^{2,1}+h_{\widetilde{\cal M}}^1 (\text{End}(\widetilde{\cal V})) &=  3 + 51 + 156 = 210\,.
}
Comparison to  the data \eqref{eq_match chiral spectrum example 1} yields
that the number of chiral zero modes did not change and, even
though the individual Hodge numbers  changed, the total number of
first order  deformations stayed the same.

This was just one possible  choice of a pair of $F$'s and $G$'s, but  actually
not the only one. We could for instance exchange a different pair, which 
involves a redefinition of $\Lambda^1$ and $\Lambda^2$ rather than $\Lambda^3$
and $\Lambda^4$. In this case we finally obtain the GLSM
\begin{equation}
\begin{aligned}
&\begin{array}{|c||c|}
\hline
x_i & \Gamma^j \\
\noalign{\hrule height 1pt}
\begin{array}{cccccccccc}
 0 & 0 & 0 & 0 & 0 & 0 & 0 & 0 & 1 & 1 \\
 0 & 0 & 0 & 0 & 1 & 1 & 1 & 1 & 0 & 0 \\
 1 & 1 & 1 & 1 & 2 & 2 & 2 & 0 & 0 & 0
\end{array}
&
\begin{array}{cccc}
 -1 & -0 & -0 & -1 \\
 -1 & -2 & -1 & -0 \\
 -2 & -4 & -3 & -1
\end{array}\\
\hline
\end{array}
\\[0.1cm]
&\begin{array}{|c||c|}
\hline
\Lambda^a & p_l \\
\noalign{\hrule height 1pt}
\begin{array}{cccccccc}
 1 & 0 & 0 & 0 & 0 & 0 & 0 & 0 \\
 0 & 1 & 1 & 1 & 0 & 0 & 0 & 1 \\
 0 & 3 & 2 & 2 & 1 & 1 & 1 & 0
\end{array}
&
\begin{array}{ccc}
 -1 & -0 & -0 \\
 -1 & -2 & -1 \\
 -3 & -4 & -3
\end{array}\\
\hline
\end{array}
\end{aligned}
\end{equation}
and find for the massless spectrum
\eq{
h_{\widehat{\cal M}}^\bullet(\widehat{\cal V}) &= ( 0, 68 , 2, 0) \,,\\
 h_{\widehat{\cal M}}^{1,1}+h_{\widehat{\cal M}}^{2,1}+h_{\widehat{\cal M}}^1 (\text{End}(\widehat{\cal V})) &=  3 + 63 + 144 = 210\,.
}
%
%
\subsection{The dual base via a conifold transition}
The methods  described in section \ref{subsec_construction of dual models}
are applicable to almost any  $(0,2)$ GLSM. 
Starting with a model that admits a  $(2,2)$ locus, namely a heterotic
model  with standard embedding, we have seen that in the
dual $(0,2)$ model one has to introduce a new $\IP^1$.
This results in a base Calabi-Yau manifold $\widetilde {\cal M}$ whose K\"ahler
moduli space has a higher dimension than the original one for ${\cal M}$. 

The question now is, what the geometric relation
between the two Calabi-Yau manifolds ${\cal M}$ and $\widetilde {\cal M}$ is.
As already observed for a specific example  in  \cite{RalphTargetSpace1,RalphTargetSpace2},
$\widetilde{\cal M}$ seems to be  connected to  ${\cal M}$ via a conifold
transition. Let us explain this for our more generic situation in more detail.

\paragraph{Standard embedding:}
Let us first consider a Calabi-Yau manifold ${\cal M}$ with
a holomorphic vector bundle ${\cal V}$ which is a deformation
of the tangent bundle $T_{\cal M}$.
As before, let $G_1,\ldots ,G_c$ be the
intersecting hypersurfaces and $x_1,...,x_d$ the homogeneous coordinates of the
ambient space. Let us pick an arbitrary hypersurface, say $G_1$, and move to
a specific region in the complex structure moduli space where we can write
this surface as a combination of polynomials of lower degree:
\begin{eqnarray}\label{eq_conifold determinant}
G_1 = x_1\,  F_1{}^1 - x_2\,  F_2{}^1=0\,,
\end{eqnarray}
where $||F_i{}^j|| = ||G_j|| - ||x_i||$. At this point in complex
structure moduli space the manifold develops a  conifold singularity.
It is well known that it can be resolved via a small resolution \cite{Rolling}. 
This is described by  introducing two new
coordinates, $y_1$ and $y_2$, parameterizing  a $\IP^1$ 
satisfying the  two  hypersurface constraints
\begin{eqnarray}
\begin{aligned}
\tilde G_1 &:=& y_1\, x_1 + y_2\, F_2{}^1=0\\
\tilde G_B &:=& y_1\, x_2  + y_2\,  F_1{}^1=0
\end{aligned} 
\end{eqnarray}
which  can be written as\footnote{In the literature, often
the  resolution $M\to M^T$ is considered.}
\begin{eqnarray} 
\begin{aligned}
M\cdot \left(\!\begin{array}{c} y_1 \\y_2 \end{array}\!\right) = 0\,,\qquad
M:=\left(\!\begin{array}{cc} x_1 & F_2{}^1 \\ x_2 & F_1{}^1\end{array}\!\right)\; .
\end{aligned}
\end{eqnarray}
Since  $y_1$ and $y_2$ are not allowed
to vanish simultaneously, the conifold
\eqref{eq_conifold determinant} is recovered 
from $\text{det}(M) = 0$.  This is the locus to which the
resolved space degenerates
in the limit of vanishing size of the $\IP^1$.
For the degrees of the two new hypersurfaces one obtains
\begin{eqnarray}
  ||\tilde G_1|| &=& \left(\begin{array}{c} 1
             \\||G_1||-||x_2|| \end{array}\right) = \left(\begin{array}{c} 1
             \\||F_2{}^1||\end{array}\right)\; ,\\
             ||\tilde G_B|| &=& \left(\begin{array}{c} 1 \\||G_1||-||x_1|| \end{array}\right) = \left(\begin{array}{c} 1 \\||F_1{}^1||\end{array}\right)\,.
\end{eqnarray}
For the  $(0,2)$ model, where ${\cal V}$ is a deformation of the tangent bundle
$T_{\cal M}$, the degree of the   $F_1{}^1, F_2{}^1$ in
\eqref{eq_conifold determinant} is equal to the degree of the 
$F_a{}^l$ in the monad \eqref{eq_general monad}.
Therefore the conifold  transition  is equivalent to the transformation
of the base described in \ref{subsec_construction of dual models}.
Thus, for the target space dual pair
\eq{
        ({\cal M}, {\cal V}) \longleftrightarrow (\widetilde{\cal M}, \widetilde{\cal V})
}
where ${\cal V}$ a deformation of $T_{\cal M}$, the two base manifolds
${\cal M}$ and $\widetilde{\cal M}$ are connected via a conifold transition.
In contrast to the conifold transition for type II superstrings, i.e.~for $(2,2)$ models,
here $\widetilde {\cal V}\ne T_{\widetilde{\cal M}}$.  Moreover,  we are not 
claiming that there is a physically smooth transition between the 
two configurations. 
On the contrary, our point is that the two $(0,2)$ models
are isomorphic descriptions of the same stringy geometry.

As an example from \cite{Rolling}, consider the Calabi-Yau manifold given by
the complete intersection of two surfaces of degree 4 and 2 in $\IP^6$,
i.e.~${\cal M}=\IP^6[4,2]$. Here we have
\eq{
\label{eq_match chiral spectrum 4}
 h_{{\cal M}}^\bullet(T_{\cal M}) &= ( 0, 89,1,0)\,\\
 h_{{\cal M}}^{1,1}+h_{{\cal M}}^{2,1}+h_{{\cal M}}^1 (\text{End}(T_{\cal M})) &=  1 + 89 + 190 = 280\,.
}
It has been shown in \cite{Rolling} that via a conifold transition, this
manifold is connected to the new Calabi-Yau $\widetilde{\cal M}$ defined as
$$
\begin{array}{|c||c|}
\hline
x_i & \Gamma^j \\
\noalign{\hrule height 1pt}
\begin{array}{cccccccc}
 0 & 0 & 0 & 0 & 0 & 0 & 1 & 1 \\
 1 & 1 & 1 & 1 & 1 & 1 & 0 & 0
\end{array}
&
\begin{array}{ccc}
 -0 & -1 & -1 \\
 -4 & -1 & -1
\end{array}\\
\hline
\end{array}\,,
$$
which has Hodge numbers $(h_{\widetilde{\cal M}}^{2,1},h_{\widetilde{\cal M}}^{2,1})=(86,2)$ 
and $h^1_{\widetilde{\cal M}}({\rm End}(T_{\widetilde{\cal M}}))=188$
bundle moduli, which gives a total number of  276 massless singlets
for the corresponding $(2,2)$ model. In our situation, the dual vector
bundle is different and given by
$$
\begin{array}{|c||c|}
\hline
\Lambda^a & p_l \\
\noalign{\hrule height 1pt}
\begin{array}{cccccc}
 0 & 0 & 0 & 0 & 1 & 0 \\
 1 & 1 & 1 & 1 & 0 & 2
\end{array}
&
\begin{array}{cc}
 -0 & -1 \\
 -4 & -2
\end{array}\\
\hline
\end{array}\\\; .
$$
We compute 
\eq{
 h_{\widetilde{\cal M}}^\bullet(\widetilde{\cal V}) &= (0 , 89 , 1, 0)\,,\\
 h_{\widetilde{\cal M}}^{1,1}+h_{\widetilde{\cal M}}^{2,1}+h_{\widetilde{\cal M}}^1 (\text{End}(\widetilde{\cal V})) &=  2 + 86 + 192 = 280\; 
}
so that  the number  of  complex structure moduli decreased by three, 
whereas the number of bundle and K\"ahler moduli increased
by two and one, respectively.

\paragraph{Generic case:}
Turning now to the case of a non-standard embedding, the story
changes only slightly.   
If we can find a point in the complex structure and bundle moduli space such
that two of the $F$'s actually appear in one and the same $G$
as\footnote{This will
always happen as long as the degree of the $F$'s are less or
equal to the degree of the $G$.}
\begin{equation}\label{eq_conifold determinant 2}
G_1 = U_1\,  F_{1}{}^{1} - U_2\,  F_{2}{}^{1}=0\,,
\end{equation}
where the  $U_i$ are  homogeneous polynomials such that
\eq{
||U_i|| = S_1-M_{1}+N_i, \qquad {\rm for}\ i=1,2\; .
}
As before, \eqref{eq_conifold determinant 2} defines a conifold
singularity, which can be resolved by blowing up $\IP^1$s over
the nodal points.
This is described by introducing two new coordinates $y_1,y_2$
parameterizing the $\IP^1$ and the two hypersurfaces
\eq{
\tilde G_1 &:= y_1\, U_1 + y_2\,  F_{2}{}^{1}=0\\
\tilde G_B &:= y_1\, U_2 + y_2\,  F_{1}{}^{1}=0\; ,
}
which can also be written as
\eq{
M\cdot \left(\!\begin{array}{c} y_1 \\y_2 \end{array}\!\right) = 0\,,\qquad
M:=\left(\!\begin{array}{cc} U_1 & F_{2}{}^{1} \\ U_2 &
  F_{1}{}^{1}\end{array}\!\right)\; .
}
The new degrees of the new coordinates and constraints are given as
\begin{eqnarray}
\begin{aligned}
&||y_2|| &=&\left(\!\begin{array}{c} 1 \\\vec 0\end{array}\!\right)\; ,\qquad &&\text{} \qquad
             ||y_1|| = \left(\!\begin{array}{c} 1
             \\||F_{1}{}^{1}||+||F_{2}{}^{1}||-S_1\end{array}\!\right)\; , \\[0.2cm]
 &||\tilde G_1|| &=&\left(\!\begin{array}{c} 1
             \\||F_{2}{}^{1}||\end{array}\!\right)\; ,\qquad &&\text{} \qquad
             ||\tilde G_B|| = \left(\!\begin{array}{c} 1
               \\||F_{1}{}^{1}||\!\end{array}\right) \; .
\end{aligned}
\end{eqnarray}
This is precisely what we obtained for the dual Calabi-Yau manifold
$\widetilde{\cal M}$ in subsection \ref{subsec_construction of dual models}.
Therefore, also for  this more  generic case the two base manifolds
are connected by a conifold transition. 

\subsection{Chains of dual models}\label{subsec_chains of dual models}
As we have explained in the beginning of this section, the proposed
construction of potentially dual models is pretty independent of the choice of
the monad, unless the data is chosen so badly, that no exchange satisfying
\eqref{eq_F and G condition} is possible. 
Therefore, one is  free to iterate the procedure
to produce dual $(0,2)$ models,  until one arrives at  a
monad  already obtained before. Depending on the initial GLSM data of
$({\cal M}_0,{\cal V}_0)$,
this can lead  to 
quite a number of dual configurations $({\cal M}_i,{\cal V}_i)$,
$i=1,\ldots,N$. In all cases investigated the number $N$ is
finite.

To show one example we choose a product of projective spaces, where the hypersurfaces have multi-degrees containing only 1's or 0's. The starting point is given by the $(2,2)$ model 
$$
\begin{array}{|c||c||c||c|}
\hline
x_i & -\Gamma^j & \Lambda^a & -p_l \\
\noalign{\hrule height 1pt}
\begin{array}{c}
\IP^2\\
\IP^2\\
\IP^4
\end{array}
&
\begin{array}{ccccc}
 0 & 0 & 1 & 1 & 1 \\
 1 & 1 & 1 & 0 & 0 \\
 1 & 1 & 1 & 1 & 1
\end{array}
&
\begin{array}{ccccccccccc}
 0 & 0 & 0 & 0 & 0 & 0 & 1 & 0 & 1 & 0 & 1 \\
 0 & 1 & 0 & 1 & 0 & 1 & 0 & 0 & 0 & 0 & 0 \\
 1 & 0 & 1 & 0 & 1 & 0 & 0 & 1 & 0 & 1 & 0
\end{array}
&
\begin{array}{ccccc}
 0 & 0 & 1 & 1 & 1 \\
 1 & 1 & 1 & 0 & 0 \\
 1 & 1 & 1 & 1 & 1
\end{array}\\
\hline
\end{array}\; ,
$$
where the first column is meant to be $\IP^2\times\IP^2\times\IP^4$. The topological data of this configuration is given by
\eq{
 h_{{\cal M}}^\bullet({\cal V}) &= (0,44,3,0)\,,\\
 h_{{\cal M}}^{1,1}+h_{{\cal M}}^{2,1}+h_{{\cal M}}^1 ({\rm End}({\cal V})) &=  3 + 44 + 48 = 95\,.
}
For this example we can apply the procedure five times until we do not obtain
anything new. All these new monads are topologically different and all have
the same chiral spectrum as the initial one:
$$
\begin{array}{|c||c|c||c|c|c|c|c}
\hline
\text{Nr.} & h_{{\cal M}}^1({\cal V}) & h_{{\cal M}}^2({\cal V}) & h_{{\cal
    M}}^{1,1} & h_{{\cal M}}^{2,1} & h_{{\cal M}}^1(\text{End}({\cal V})) &
     {\rm D}({\cal M},{\cal V}) \\
\noalign{\hrule height 1pt}
1 & 44 & 3 & 4 & 42 & 49 & 95 \\\hline
2 & 44 & 3 & 5 & 40 & 50 & 95 \\\hline
3 & 44 & 3 & 6 & 38 & 51 & 95 \\\hline
4 & 44 & 3 & 7 & 36 & 52 & 95  \\
\hline
\end{array}
$$
The defining data can be derived to be
\paragraph*{Model 0:}
\footnotesize{
\begin{eqnarray*}
&
\begin{array}{|c||c||c||c|}
\hline
\begin{array}{c}
\IP^2\\
\IP^2\\
\IP^4
\end{array}
&
\begin{array}{ccccc}
 0 & 0 & 1 & 1 & 1 \\
 1 & 1 & 1 & 0 & 0 \\
 1 & 1 & 1 & 1 & 1
\end{array}
&
\begin{array}{ccccccccccc}
 0 & 0 & 0 & 0 & 0 & 0 & 0 & 0 & 1 & 1 & 1 \\
 0 & 0 & 0 & 0 & 0 & 1 & 1 & 1 & 0 & 0 & 0 \\
 1 & 1 & 1 & 1 & 1 & 0 & 0 & 0 & 0 & 0 & 0
\end{array}
&
\begin{array}{ccccc}
 0 & 0 & 1 & 1 & 1 \\
 1 & 1 & 1 & 0 & 0 \\
 1 & 1 & 1 & 1 & 1
\end{array}\\
\hline
\end{array}
\end{eqnarray*}
}

\paragraph*{Model 1:}
\footnotesize{
\begin{eqnarray*}
&
\begin{array}{|c||c||c||c|}
\hline
\begin{array}{c}
\IP^1\\
\IP^2\\
\IP^2\\
\IP^4
\end{array}
&
\begin{array}{cccccc}
 1 & 0 & 0 & 0 & 0 & 1 \\
 0 & 0 & 1 & 1 & 1 & 0 \\
 1 & 1 & 1 & 0 & 0 & 0 \\
 0 & 1 & 1 & 1 & 1 & 1
\end{array}
&
\begin{array}{ccccccccccc}
 1 & 0 & 0 & 0 & 0 & 0 & 0 & 0 & 0 & 0 & 0 \\
 0 & 0 & 0 & 0 & 0 & 0 & 0 & 0 & 1 & 1 & 1 \\
 0 & 0 & 0 & 0 & 0 & 1 & 1 & 1 & 0 & 0 & 0 \\
 0 & 1 & 1 & 1 & 1 & 1 & 0 & 0 & 0 & 0 & 0
\end{array}
&
\begin{array}{ccccc}
 1 & 0 & 0 & 0 & 0 \\
 0 & 0 & 1 & 1 & 1 \\
 1 & 1 & 1 & 0 & 0 \\
 1 & 1 & 1 & 1 & 1
\end{array}\\
\hline
\end{array}
\end{eqnarray*}
}

\paragraph*{Model 2:}

\footnotesize{
\begin{eqnarray*}
&
\begin{array}{|c||c||c||c|}
\hline
\begin{array}{c}
\IP^1\\
\IP^1\\
\IP^2\\
\IP^2\\
\IP^4
\end{array}
&
\begin{array}{ccccccc}
 0 & 1 & 0 & 0 & 0 & 0 & 1 \\
 1 & 0 & 0 & 0 & 0 & 1 & 0 \\
 0 & 0 & 1 & 1 & 1 & 0 & 0 \\
 1 & 1 & 1 & 0 & 0 & 0 & 0 \\
 0 & 0 & 1 & 1 & 1 & 1 & 1
\end{array}
&
\begin{array}{ccccccccccc}
 0 & 1 & 0 & 0 & 0 & 0 & 0 & 0 & 0 & 0 & 0 \\
 1 & 0 & 0 & 0 & 0 & 0 & 0 & 0 & 0 & 0 & 0 \\
 0 & 0 & 0 & 0 & 0 & 0 & 0 & 0 & 1 & 1 & 1 \\
 0 & 0 & 0 & 0 & 0 & 1 & 1 & 1 & 0 & 0 & 0 \\
 0 & 0 & 1 & 1 & 1 & 1 & 1 & 0 & 0 & 0 & 0
\end{array}
&
\begin{array}{ccccc}
 0 & 1 & 0 & 0 & 0 \\
 1 & 0 & 0 & 0 & 0 \\
 0 & 0 & 1 & 1 & 1 \\
 1 & 1 & 1 & 0 & 0 \\
 1 & 1 & 1 & 1 & 1
\end{array}\\
\hline
\end{array}
\end{eqnarray*}
}

\paragraph*{Model 3:}
\footnotesize{
\begin{eqnarray*}
&
\begin{array}{|c||c||c||c|}
\hline
\begin{array}{c}
\IP^1\\
\IP^1\\
\IP^1\\
\IP^2\\
\IP^2\\
\IP^4
\end{array}
&
\begin{array}{cccccccc}
 0 & 0 & 0 & 1 & 0 & 0 & 0 & 1 \\
 0 & 1 & 0 & 0 & 0 & 0 & 1 & 0 \\
 1 & 0 & 0 & 0 & 0 & 1 & 0 & 0 \\
 0 & 0 & 1 & 1 & 1 & 0 & 0 & 0 \\
 1 & 1 & 1 & 0 & 0 & 0 & 0 & 0 \\
 0 & 0 & 1 & 0 & 1 & 1 & 1 & 1
\end{array}
&
\begin{array}{ccccccccccc}
 0 & 0 & 0 & 1 & 0 & 0 & 0 & 0 & 0 & 0 & 0 \\
 0 & 1 & 0 & 0 & 0 & 0 & 0 & 0 & 0 & 0 & 0 \\
 1 & 0 & 0 & 0 & 0 & 0 & 0 & 0 & 0 & 0 & 0 \\
 0 & 0 & 0 & 0 & 0 & 0 & 0 & 0 & 1 & 1 & 1 \\
 0 & 0 & 0 & 0 & 0 & 1 & 1 & 1 & 0 & 0 & 0 \\
 0 & 0 & 1 & 0 & 1 & 1 & 1 & 0 & 0 & 1 & 0
\end{array}
&
\begin{array}{ccccc}
 0 & 0 & 0 & 1 & 0 \\
 0 & 1 & 0 & 0 & 0 \\
 1 & 0 & 0 & 0 & 0 \\
 0 & 0 & 1 & 1 & 1 \\
 1 & 1 & 1 & 0 & 0 \\
 1 & 1 & 1 & 1 & 1
\end{array}\\
\hline
\end{array}
\end{eqnarray*}
}

\paragraph*{Model 4:}
\footnotesize{
\begin{eqnarray*}
&
\begin{array}{|c||c||c||c|}
\hline
\begin{array}{c}
\IP^1\\
\IP^1\\
\IP^1\\
\IP^1\\
\IP^2\\
\IP^2\\
\IP^4
\end{array}
&
\begin{array}{ccccccccc}
 0 & 0 & 0 & 0 & 1 & 0 & 0 & 0 & 1 \\
 0 & 0 & 0 & 1 & 0 & 0 & 0 & 1 & 0 \\
 0 & 1 & 0 & 0 & 0 & 0 & 1 & 0 & 0 \\
 1 & 0 & 0 & 0 & 0 & 1 & 0 & 0 & 0 \\
 0 & 0 & 1 & 1 & 1 & 0 & 0 & 0 & 0 \\
 1 & 1 & 1 & 0 & 0 & 0 & 0 & 0 & 0 \\
 0 & 0 & 1 & 0 & 0 & 1 & 1 & 1 & 1
\end{array}
&
\begin{array}{ccccccccccc}
 0 & 0 & 0 & 0 & 1 & 0 & 0 & 0 & 0 & 0 & 0 \\
 0 & 0 & 0 & 1 & 0 & 0 & 0 & 0 & 0 & 0 & 0 \\
 0 & 1 & 0 & 0 & 0 & 0 & 0 & 0 & 0 & 0 & 0 \\
 1 & 0 & 0 & 0 & 0 & 0 & 0 & 0 & 0 & 0 & 0 \\
 0 & 0 & 0 & 0 & 0 & 0 & 0 & 0 & 1 & 1 & 1 \\
 0 & 0 & 0 & 0 & 0 & 1 & 1 & 1 & 0 & 0 & 0 \\
 0 & 0 & 1 & 0 & 0 & 1 & 1 & 0 & 0 & 1 & 1
\end{array}
&
\begin{array}{ccccc}
 0 & 0 & 0 & 0 & 1 \\
 0 & 0 & 0 & 1 & 0 \\
 0 & 1 & 0 & 0 & 0 \\
 1 & 0 & 0 & 0 & 0 \\
 0 & 0 & 1 & 1 & 1 \\
 1 & 1 & 1 & 0 & 0 \\
 1 & 1 & 1 & 1 & 1
\end{array}\\
\hline
\end{array}
\end{eqnarray*}
}
\normalsize
\section{Landscape studies}\label{sec_landscape studies}
So far, we have verified the proposed
general target space duality between $(0,2)$ GLSMs
only for a couple of examples.
In fact, invoking a fast computer
implementation, we have actually performed a large scale landscape study
of this target space duality. We generated  ten-thousands of 
candidate dual models and then computed the massless particle
spectra, i.e.~the number of chiral matter fields
and the number of massless gauge singlets  ${\rm D}({\cal M},{\cal V})$.  
Let us report on our findings.

\subsection{The scanning algorithm}

The algorithm to generate dual $(0,2)$ GLSMs enabled us to 
perform a scan over many different models. While one performs the duality
transformation it might happen that new singularities arise and in general it
may be hard to resolve them properly. For that reason, we only considered those
cases where almost no new singularities appeared. Our scanning algorithm looks
as follows, starting with 
step 1:
\small{
\begin{equation*}
\xymatrix{
  \boxed{\parbox{2.3cm}{\center \vspace{-0.5cm} {\bfseries Step 1:}\\ Go to next model in list}} \ar[r]& 
 \boxed{\parbox{2.2cm}{\center \vspace{-0.5cm} {\bfseries Step 2:}\\ Triangulize polytope via TOPCOM}}\ar[r] &
 \boxed{\parbox{2.2cm}{\center \vspace{-0.5cm}   {\bfseries Step 3:}\\ Generate SR ideal, inters. numbers via Schubert}}\ar[r] & 
 \boxed{\parbox{2.2cm}{\center \vspace{-0.5cm} {\bfseries Step 4:}\\ Calculate line bundles from Euler and monad complex}} \ar[dl]
 \\
 \boxed{\parbox{2.2cm}{\center \vspace{-0.5cm} {\bfseries Step 7:}\\ Calculate
     all $h^1_{\cal M}(\text{End}({\cal V}))$}} \ar[u]& 
 \boxed{\parbox{2.2cm}{\center \vspace{-0.5cm} {\bfseries Step 6:}\\ Generate GLSM data of next configuration}} \ar[u]^{\text{if possible}}\ar[l]^{\text{if not}}& 
 \boxed{\parbox{2.3cm}{\center \vspace{-0.5cm}   {\bfseries Step 5:}\\ Compare $\sum_{i=1}^3(-)^ih^i$ to holom. $\chi$}} \ar[l]^{\text{agree}}\ar[r]^{\text{don't}}_{\text{agree}}&
 \boxed{\parbox{2.2cm}{\center \vspace{-0.5cm} Delete configuration}}
 \\}
\end{equation*}
}
\normalsize
We ran through two different lists (mentioned in step 1). The first one
contained Calabi-Yau manifolds defined via single hypersurfaces in 
toric varieties. We took
the ambient spaces out of the list from \cite{Kreuzer:2000xy} available on the
website of Maximilian Kreuzer \cite{KreuzerList} and the second list contains
codimension 2 complete intersections in weighed projective spaces which is
part of the list presented in \cite{Klemm:2004km} and available at
\cite{KlemmList}. To resolve the ambient spaces and also to generate the set
of nef partitions to obtain the codimension 2 Calabi-Yaus, we used PALP
\cite{PALP}. For the remaining steps several packages as TOPCOM \cite{TOPCOM}
Schubert \cite{Schubert} and of course
\cohomCalgKoszul~\cite{cohomCalg:Implementation} along with some Mathematica
routines were employed. For the interplay of TOPCOM and Schubert we use the
(not published) Toric Triangulizer \cite{ToricTriangulizer}.

\subsection{Hypersurfaces in toric varieties}
\begin{figure}[htbp]
  \subfigure[Hypersurface models part 1]{
	  \includegraphics[angle=0,scale=0.6825]{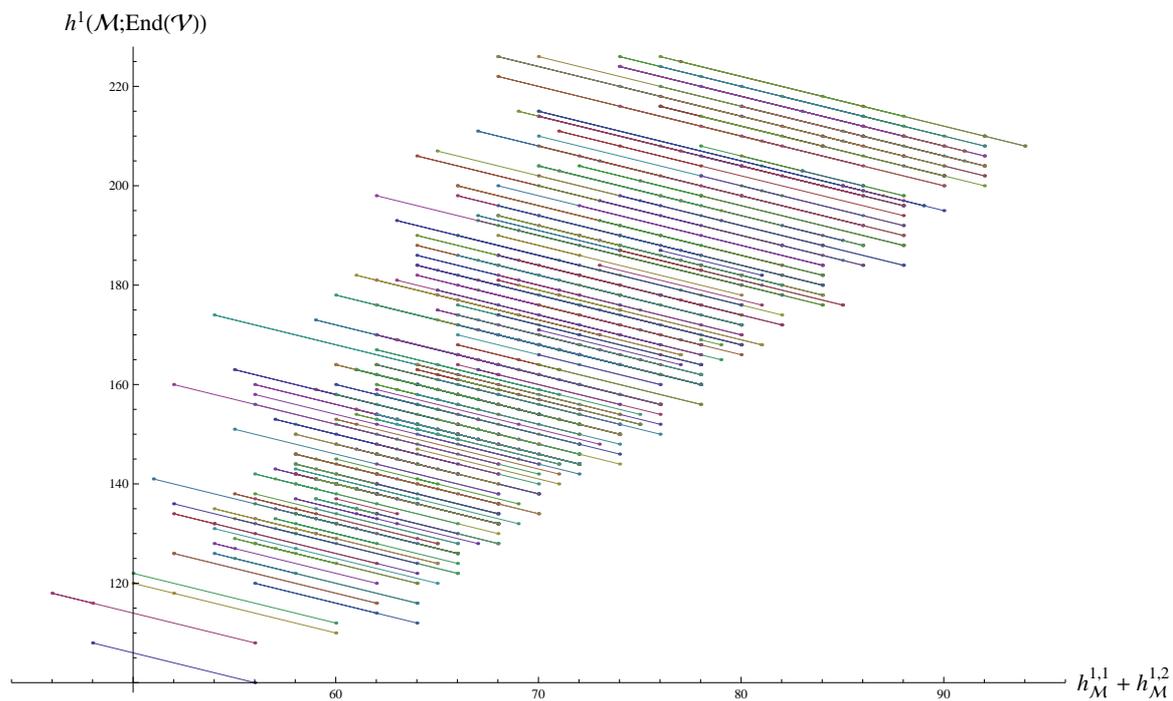}
  }
  \subfigure[Hypersurface models part 2]{
	 \includegraphics[angle=0,scale=0.6825]{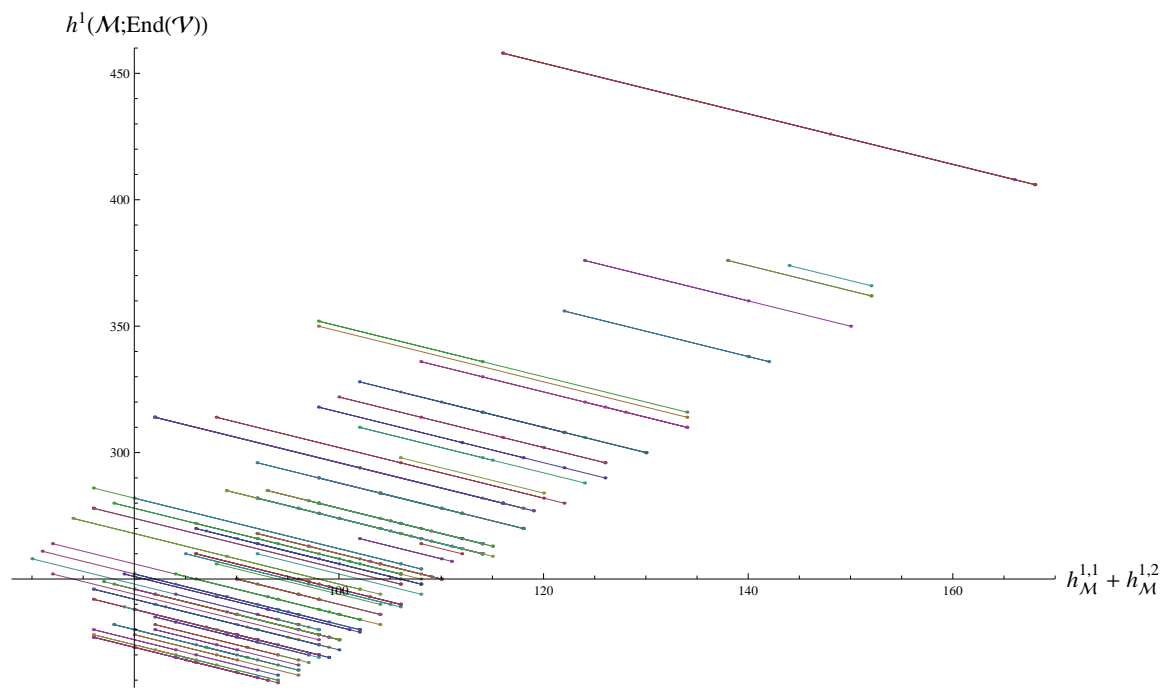}
  }
\caption{Plot of the topological data of hypersurfaces in toric varieties and their codimension two duals with full agreement. Each line corresponds to one class of dual models. Different colored overlapping lines correspond to different classes.}
\label{fig_plot hypersurfaces}
\end{figure}

Our first scan ran over the list of hypersurfaces in toric varieties
\cite{KreuzerList} where we considered all toric varieties with 7, 8 and 9
lattice points which make altogether $1,085$. Starting from this geometry, we
performed all first duals to each of those models in the way described in
\ref{subsec_construction of dual models} where we always introduced exactly
one new hypersurface. Hence the dual models of each hypersurface Calabi-Yau
are here codimension 2 complete intersections in toric varieties. Since
already many of the duals are obtained by only performing the duality
procedure once, we did not perform duals of duals as shown in
\ref{subsec_chains of dual models}. In figure \ref{fig_plot hypersurfaces} we
displayed  all models with full agreement of the chiral spectrum and the sum of
complex structure, K\"ahler and bundle deformations, i.e.~${\rm D}({\cal
  M},{\cal V})$. Some details on the full
analysis are shown in table \ref{tab_results hypersurfaces}.
\begin{table}[h]
\begin{tabular}{ccccccc}
\parbox{2cm}{\center Different classes}&
\parbox{2cm}{\center Possibly smooth models} & 
\parbox{2cm}{\center Classes without duals}&
\parbox{2cm}{\center Models with matching spectrum}&
\parbox{2cm}{\center Models with full agreement}&
\parbox{2cm}{\center Computed (different) line bundle cohom.}\\
\hline
1,085 & 4,507 & 42 & \parbox{2cm}{\center 4,144 (100\%) } & \parbox{2cm}{\center 1509 (94.6\%)} & \parbox{2cm}{\center (1,481,539) 3,069,067}\\

\end{tabular}
\caption{Some data on the landscape study: Starting point are hypersurfaces in
  toric varieties that are given by the polytopes with at most 9 lattice
  points. The percent numbers in the parentheses in column 4 and 5 only cover
  models where these numbers could actually be calculated. In column 5, by ``full agreement'' we mean that the chiral spectrum of dual models as well as the sum of complex structure, K\"ahler and bundle deformations agree.}
\label{tab_results hypersurfaces}
\end{table}

\subsection{CICY of two hypersurfaces}
\begin{figure}[htbp]
  \subfigure[Complete intersection models part 1]{
	  \includegraphics[angle=0,scale=0.6825]{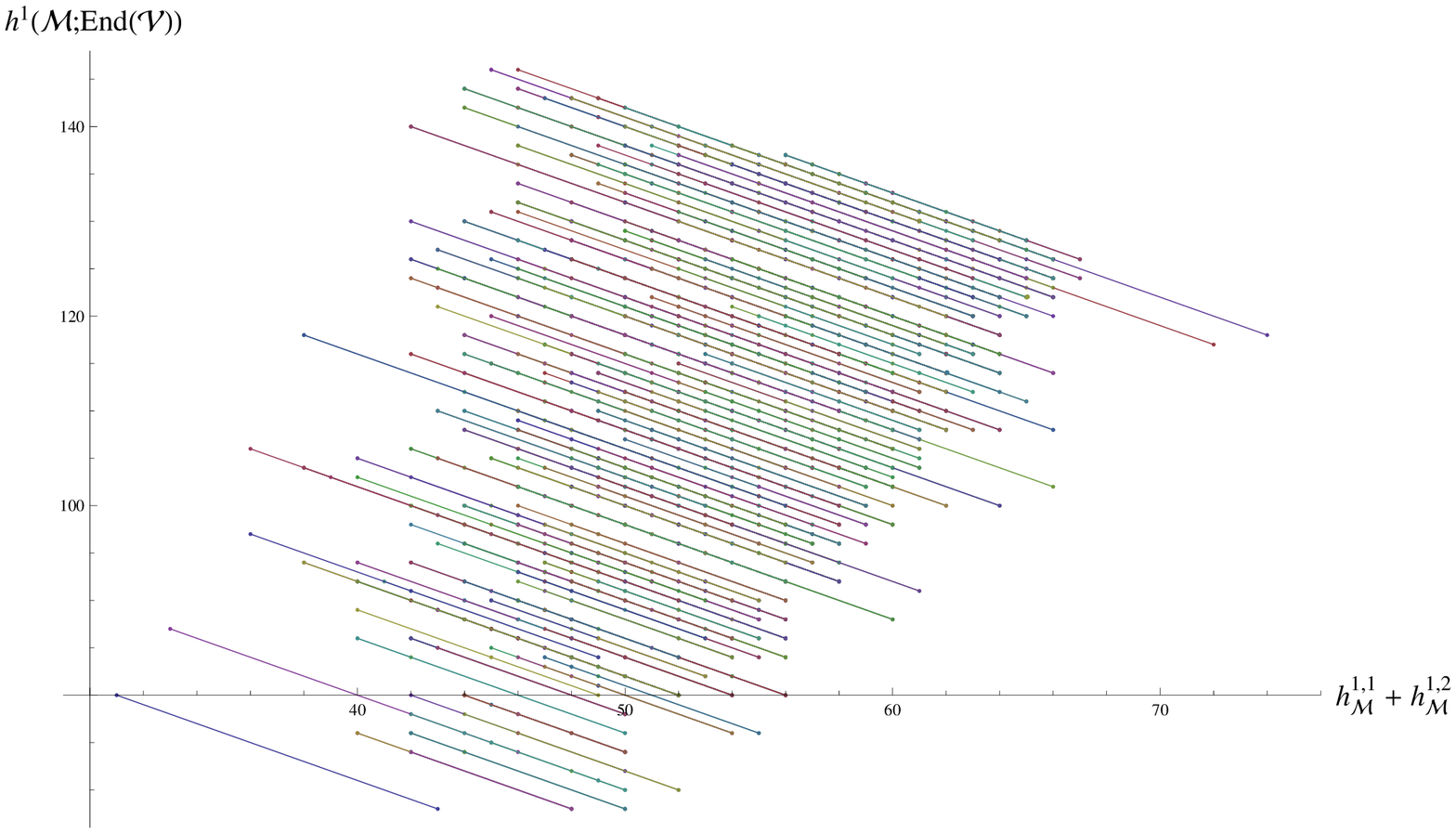}
  }
  \subfigure[Complete intersection models part 2]{
	 \includegraphics[angle=0,scale=0.6825]{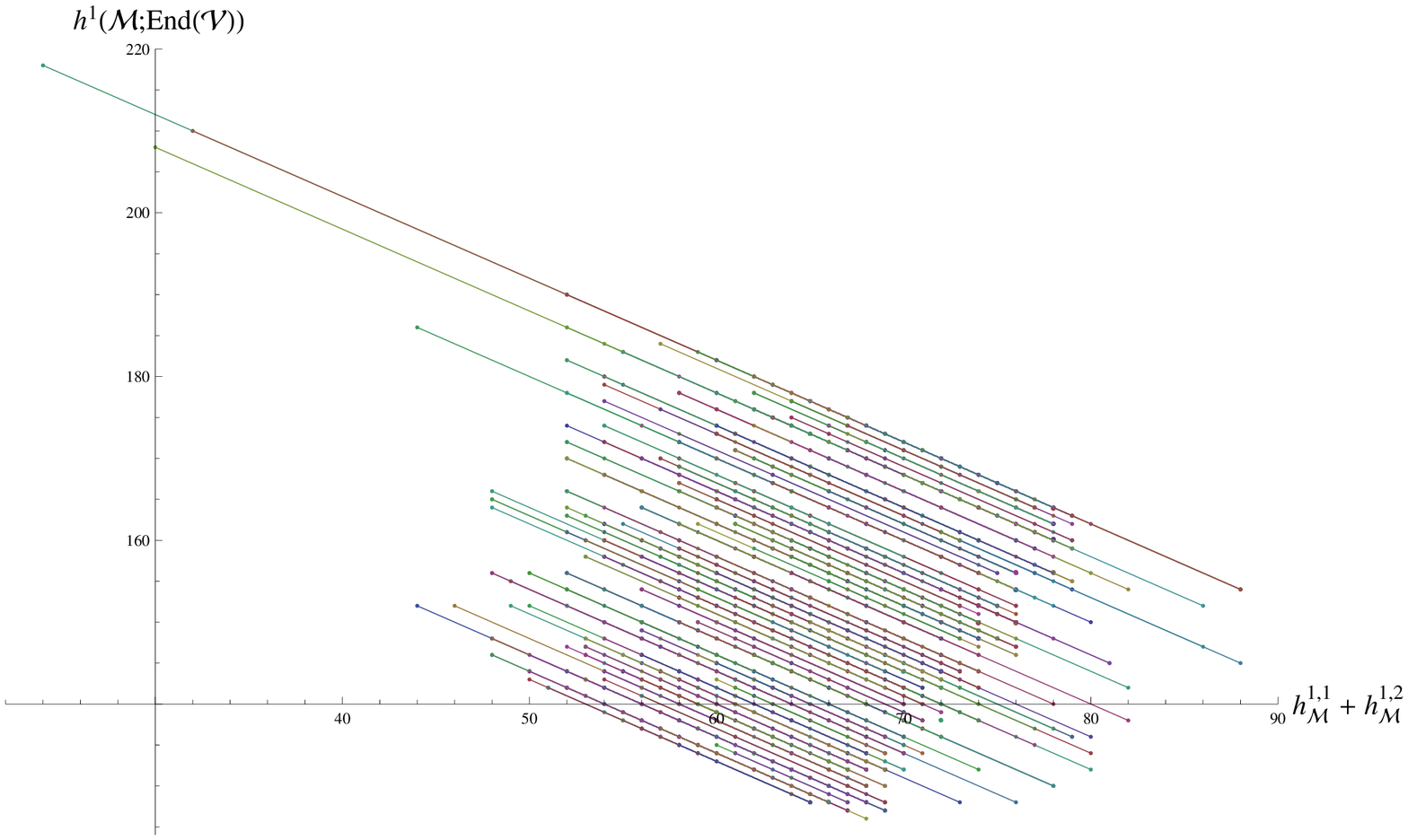}
  }
\caption{Plot of the topological data of codimension two complete intersections in weighted projective spaces and their codimension three duals. Each line corresponds to one class of dual models. Different colored overlapping lines correspond to different classes.}
\label{fig_plot CICY 1-2}
\end{figure}

\begin{figure}[htbp]
  \subfigure[Complete intersection models part 3]{
	  \includegraphics[angle=0,scale=0.6825]{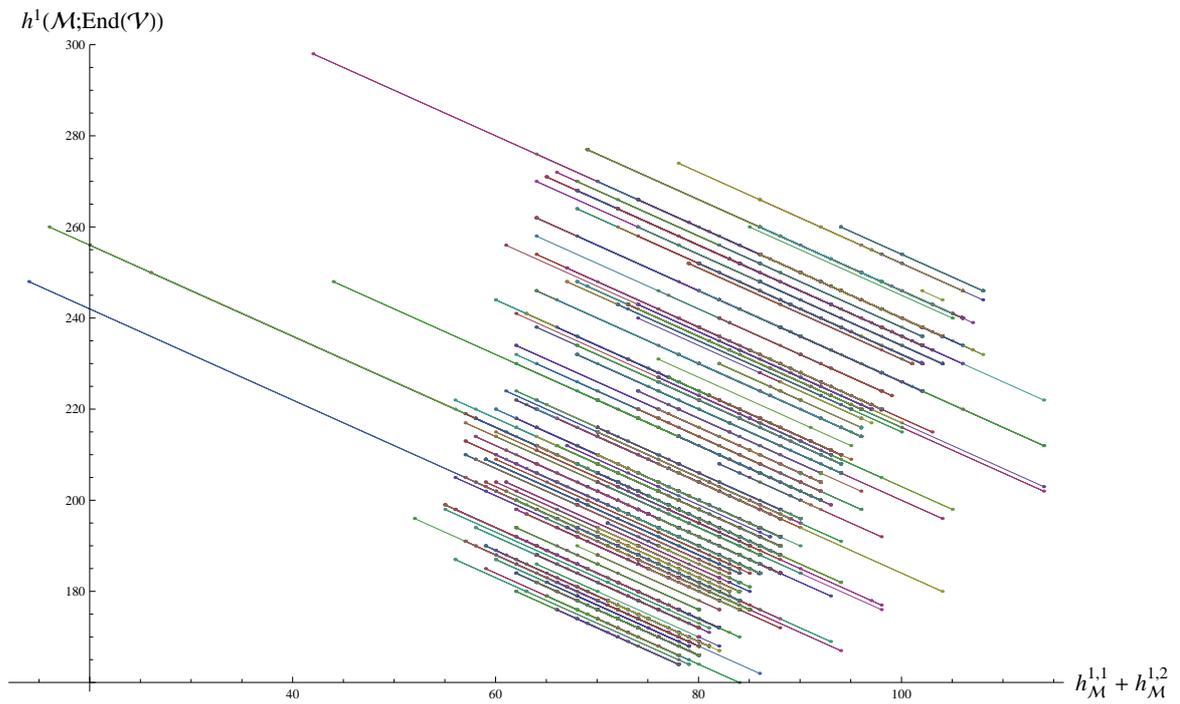}
  }
  \subfigure[Complete intersection models part 4]{
	 \includegraphics[angle=0,scale=0.6825]{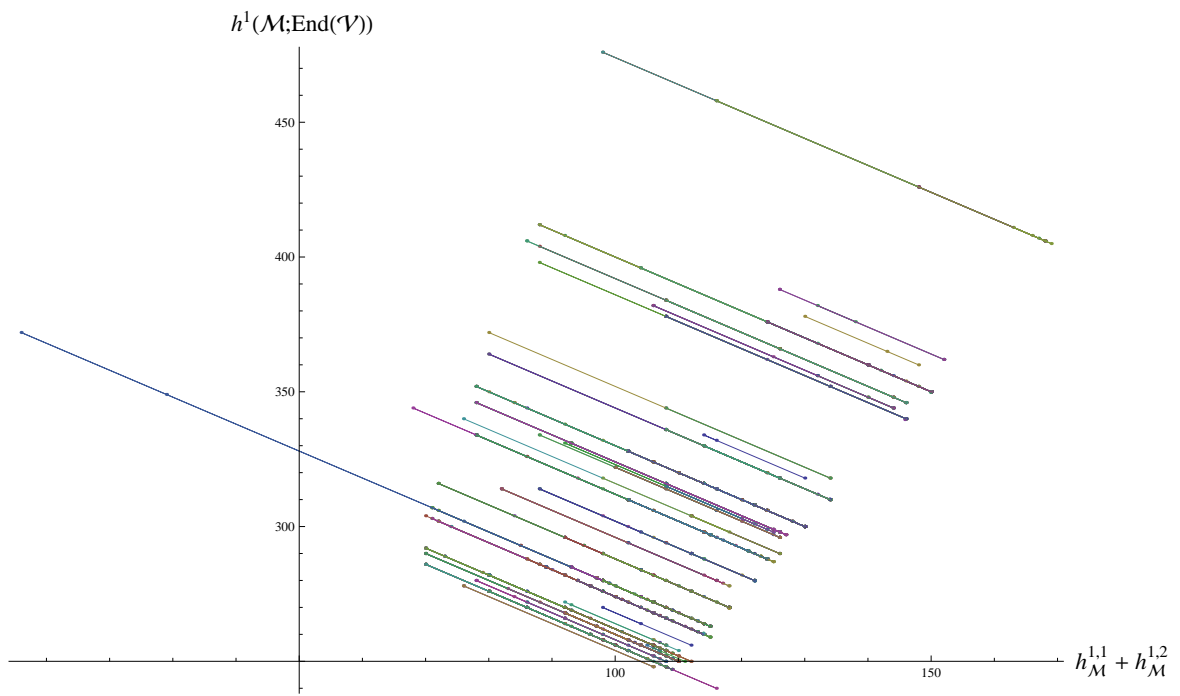}
  }
\caption{Plot of the topological data of codimension two complete intersections in weighted projective spaces and their codimension three duals. Each line corresponds to one class of dual models. Different colored overlapping lines correspond to different classes.}
\label{fig_plot CICY 3-4}
\end{figure}

As a second scan we took a list of codimension 2 complete intersections in
weighted projective spaces as a start, rather than just single hypersurfaces. 
This
list can be found online at \cite{KlemmList}. For our scan we simply ran
through the first $2,780$ ambient spaces and chose the $16,029$
possible nef partitions as starting points. All these nef partitions
correspond to  topologically distinct Calabi-Yau manifolds that are complete
intersections of two hypersurfaces in the corresponding weighted projective
space. All dual models  are codimension three complete intersections in toric varieties. In figures
\ref{fig_plot CICY 1-2} and \ref{fig_plot CICY 3-4} we have displayed all  the
models where a full agreement of deformations and chiral spectrum was
found. In table \ref{tab_results complete intersections} we provide the
summary of some details on the full scan.
\begin{table}[h]
\begin{tabular}{ccccccc}
\parbox{2.0cm}{\center Different classes}&
\parbox{2cm}{\center Possibly smooth models} & 
\parbox{2cm}{\center Classes without duals}&
\parbox{2cm}{\center Models with matching spectrum}&
\parbox{2cm}{\center Models with full agreement}&
\parbox{2cm}{\center Computed (different) line bundle cohom.}\\
\hline
16,961 & 79,204 & 718 & \parbox{2cm}{\center 64,332 \\ (85 \%) } & \parbox{2cm}{\center 20,336 (91\%)} & \parbox{2cm}{\center (38,807,002) 109,228,732}\\
\end{tabular}
\caption{Some data on landscape study. Starting point are codimension two complete intersections in weighted projective spaces.}
\label{tab_results complete intersections}
\end{table}

\subsection{The mismatch}\label{sec_The Mismatch}
While we were performing the scan over the landscape, we found that the
duality holds in most the cases, but not in all of them. In two different ways
it actually happened to fail i.e.~either such that the chiral spectrum did not
match or that that the sum of total deformations of the Calabi-Yau and the
bundle did not match. The chiral spectrum can fail to match for the following reasons:

\begin{itemize}
 \item The Calabi-Yau manifold is not smooth and still contains
   singularities\footnote{The check of the holomorphic Euler characteristic
      for line bundles over the Calabi-Yau  is only a necessary
   condition for smoothness.}. In addition, the monad might not define
   a smooth vector bundle, but for instance merely a coherent
   sheaf with non-constant  rank  (see e.g.  \cite{DistlerGreeneResolving,RalphTargetSpace2} for a correct 
     treatment of such configurations).
 \item During the scan we did not explicitly check whether the model we
   started with actually admits  a phase that allows for the redefinition of the
   corresponding fields. So it might happen that such a phase did not exist
   which would forbid the exchange of specific $F$'s and $G$'s.
\end{itemize}

\noindent
The mismatch of  
${\rm D}({\cal M},{\cal V})\ne  {\rm D}(\widetilde{\cal M},\widetilde{\cal
  V})$  can of course be traced back to the same reasons, but 
could also be happening since we assumed the bundle to be stable and furthermore the map $\varphi$ in \eqref{esiwien} to be surjective.

\paragraph{Example:}
Let us
present a simple example where a  mismatch occurs.
Consider the following configuration with standard embedding:
$$
\begin{array}{|c||c|}
\hline
x_i & \Gamma^j \\
\noalign{\hrule height 1pt}
\begin{array}{cccccccc}
 -1 & 0 & 0 & 1 & 1 & 1 \\
 1 & 1 & 1 & 0 & 0 & 0
\end{array}
&
\begin{array}{ccc}
 -2 \\
 -3
\end{array}\\
\hline
\end{array}\,.
$$
One possible dual model of this configuration can be be obtained as
\begin{equation}
\begin{aligned}
\begin{array}{|c||c|}
\hline
x_i & \Gamma^j \\
\noalign{\hrule height 1pt}
\begin{array}{cccccccccc}
 0 & 0 & 0 & 0 & 0 & 0 & 1 & 1 \\
 -1 & 0 & 0 & 1 & 1 & 1 & 0 & 0 \\
 1 & 1 & 1 & 0 & 0 & 0 & 3 & 0
\end{array}
&
\begin{array}{cccc}
 -1 & -1 \\
 -1 & -1 \\
 -3 & -3
\end{array}\\
\hline
\end{array}
&
\begin{array}{|c||c|}
\hline
\Lambda^a & p_l \\
\noalign{\hrule height 1pt}
\begin{array}{cccccccc}
 0  & 0 & 0 & 1 & 0 & 0 \\
 -1 & 0 & 0 & 0 & 2 & 1 \\
 1  & 1 & 1 & 0 & 0 & 0
\end{array}
&
\begin{array}{ccc}
 -1\\
 -2\\
 -3
\end{array}\\
\hline
\end{array}\,.
\end{aligned}
\nonumber
\end{equation}
For the initial model $({\cal M},{\cal V})$ we calculate the following data
\begin{eqnarray*} 
 h_{S}^\bullet(V) &=& (0,86,2,0)\,,\\
 h_{S}^{1,1}+h_{S}^{2,1}+h_{S}^1 (\text{End}(V)) &=&  2 + 86 + 184 =272\,,
\end{eqnarray*}
whereas for the dual model we find 
\begin{eqnarray*} 
 h_{\tilde S}^\bullet(\tilde V) &=& (0,86,2,0)\,,\\
 h_{\tilde S}^{1,1}+h_{\tilde S}^{2,1}+h_{\tilde S}^1 (\text{End}(\tilde V)) &=&  3 + 78 + 195 = 276\,,
\end{eqnarray*}
We observe  that there is a mismatch of 4 for ${\rm D}({\cal M},{\cal V})$, but
at the present state it is hard to determine the precise
origin of this mismatch.

\section{Conclusions}
In this paper we have proposed a method to construct from almost any given
$(0,2)$ heterotic model, dual models that generically 
have the same massless spectra. This procedure should work basically with all
possible structure groups $SU(3),~SU(4),~SU(5)$ for the bundle and preserves
all anomaly cancellation conditions. For the special case that Fermi superfields
become uncharged in the dual model, it was suggested that an 
additional blowup of a
$\IP^1$ has to be performed.
Furthermore it was pointed out that in these cases the duality
transformation of the base could be understood as the resolution
of a conifold singularity. 

To provide evidence for our proposal, a large number of
examples were investigated where the initial models were hypersurfaces in
toric varieties and codimension two complete intersections in weighted
projective spaces. For both types the initial configuration
was the Calabi-Yau manifold  equipped with a deformation of  its
tangent bundle of $SU(3)$ structure group. A great number of
models agreed in all instances and an interpretation of the mismatch of the
bundle deformations and an explanation of the mismatch of the chiral spectrum
was suggested. There are a couple of things that would be interesting to
investigate further:

Since it was not checked explicitly whether the bundle of a dual configuration
is indeed a stable, it would be very useful  to find a way or a
requirement for the proposed procedure that ensures  stability of the
dual bundle. 

We argued  that deformations of the complex structure and of the bundle, 
that come from  global  section of line bundles on the Calabi-Yau manifold,
are unobstructed.
A mathematically rigorous
treatment of these obstructions was recently presented in
\cite{Anderson:2010mh} and it would be interesting whether potential
target space dual models also have the same number of unobstructed
deformations.

The landscape study we performed, on the one hand, was quite extensive
but, on the other  hand,  restricted
to a particular class of models, namely those that arise from standard
embeddings and hence have $SU(3)$ structure group. Even though
for vector bundles  with other  structure groups the duality was checked 
in some cases, it would be interesting to see, if this works also
for a much larger sample of models.
In this respect, further checks would be possible,
i.e.~a matching of zero modes that live in $h^1_{\cal M}(\Lambda^2 {\cal V})$ and
$h^1({\cal M},\Lambda^2 {\cal V}^\ast)$.

Since such a large number of 83,711 models was analyzed,
we are confident that a fair ratio really defines "healthy"
configurations. Nevertheless, since only consistency checks were made in order
to detect singularities of the generated spaces, a closer analysis of the
specific configuration would be necessary in order to ensure that the base
as well as the bundle are indeed smooth. Since from the list of different Hodge numbers that were generated, there were quite some that turned
out to be actually not yet discovered combinations of $(h^{2,1},h^{1,1})$, 
the explicit analysis of these spaces is a worthwhile thing to do \cite{WorkInProgress}.

For elliptically fibered Calabi-Yau threefolds with vector bundles
defined via the spectral cover construction, it is known that
there exist a higher dimensional framework, namely F-theory on
Calabi-Yau fourfolds, in which the complex structure and bundle
deformations are unified. In fact, considering a Calabi-Yau fourfould
admitting two different K3-fibrations would also imply a duality 
between two seemingly different  heterotic $(0,2)$ models. In this respect,
it is an interesting question whether also in the present  case of
$(0,2)$ GLSMs a unified description exists, where the duality is  
manifest.


\subsection*{Acknowledgment}
We would like to thank Benjamin Jurke for useful discussions and also for providing
us with his Toric Triangulizer and related {\it Mathematica} scripts. We are also grateful for comments and
discussions with Lara Anderson, Volker Braun, James Gray, Stefan Groot Nibbelink, Ilarion Melnikov and Bernhard Wurm. We would also like to thank the Erwin Schr\"odinger International Institute for Mathematical Physics (ESI) for hospitality.

\appendix

\section{Anomaly cancellation}\label{app_anomaly cancellation}

In this appendix, we show that the dual  configuration $(\widetilde{\cal
  M},\widetilde {\cal V})$ 
satisfy the anomaly cancellation conditions \eqref{eq_anomcancel}, if
they were satisfied by the  initial one $({\cal M},{\cal V})$.  
For this purpose, let us start with a general configuration 
\begin{equation*}
V_{N_1,...,N_\delta}[M_1,...,M_\gamma]\longrightarrow \IP_{Q_1,...,Q_d}[S_1,...,S_c]\,.
\end{equation*}
that satisfies the combinatorial relations \eqref{eq_anomcancel}:
\begin{eqnarray*}
  &&\sum_{a=1}^\delta  N_a^{(\alpha)} = \sum_{l=1}^\gamma  M_l^{(\alpha)},\qquad\quad
  \sum_{i=1}^d  Q_i^{(\alpha)} = \sum_{j=1}^c  S_j^{(\alpha)} \\
  &&\sum_{l=1}^\gamma  M_l^{(\alpha)} M_l^{(\beta)} - 
  \sum_{a=1}^\delta  N_a^{(\alpha)} N_a^{(\beta)} =  \sum_{j=1}^c
  S_j^{(\alpha)} 
   S_j^{(\beta)}  - \sum_{i=1}^d Q_i^{(\alpha)} Q_i^{(\beta)} \; ,
\end{eqnarray*}
for all $\alpha, \beta=1,\ldots, r$.
We want to to show that this implies that the dual configuration,
\begin{eqnarray*}
V_{\tilde{N}_1,\tilde{N}_2,N_3,...,N_\delta}[M_1,M_2,...,M_\gamma]
\longrightarrow
\IP_{Q_1,...,Q_d,B}[\tilde{S}_1,S_3,...,S_c,\tilde B]\,,
\end{eqnarray*}
with charges given in table \ref{table_new degrees of all fields}, 
still satisfies these relations. The new fields that changed comparing to the initial model read
\eq{
\begin{array}{|c|c|c|c|c|c|c|c|c|}
\hline
y_1 & y_2 & \tilde\Gamma^1  & \tilde\Gamma^B  & \tilde F_1{}^1 & \tilde F_2{}^1 & \tilde \Lambda^1 & \tilde \Lambda^2 & p_1 \\\noalign{\hrule height 1pt}
 1  	& 1					 & -1					  		 & -1								 & 0							& 1							 & 1                & 0                & -1\\\hline
 B   	& 0					 & -(M_1 - N_1) 	 & -(M_1 - N_2)			 & S_1					  & B	 						 & M_1-S_1          & M_1 - B          & -M_1
 \\\hline
\end{array}\nonumber \;.
}

\noindent
Since  $y_1$ was chosen in a way that
\eq{
 B+S_1 = ||F_1{}^1||+||F_2{}^1|| = 2M_1-N_1-N_2
}
we get
\eq{
||\tilde F_2{}^{1}|| &= \left(\!\begin{array}{c} 1
  \\ 2M_1-N_1-N_2-S_1\end{array}\! \right)\, , \\
 ||\tilde\Lambda^2|| &= \left(\!\begin{array}{c} 1
 \\ -M_1+N_1+N_2+S_1\end{array}\!\right)\; .
}

\paragraph{Linear relations:}
Let us  refer to the $U(1)$ charges that belong to the blown up $\IP^1$ as new $U(1)$ charges. The Calabi-Yau condition, i.e.~the second equation in \eqref{eq_anomcancel} for the dual model is clear for the new $U(1)$ charges. For the other $U(1)$'s it reads
\begin{eqnarray*}
\sum_{i=1}^d Q_i^{(\alpha)} + B^{(\alpha)} &=& {\sum_{j=2}^c S_j^{(\alpha)}} +
(M_1^{(\alpha)}-N_1^{(\alpha)})\\[-0.4cm] & & \phantom{aaaaaaaaaa} +(M_1^{(\alpha)}-N_2^{(\alpha)}) \\
\Leftrightarrow\ \sum_{i=1}^d Q_i^{(\alpha)} + 2M_1^{(\alpha)}-N_1^{(\alpha)}-N_2^{(\alpha)}-S_1^{(\alpha)} &=& \sum_{j=2}^c S_j^{(\alpha)}+2M_1^{(\alpha)}-N_1^{(\alpha)}-N_2^{(\alpha)} \\
\Leftrightarrow\ {\sum_{i=1}^d Q_i^{(\alpha)}} + 2M_1^{(\alpha)}-N_1^{(\alpha)}-N_2^{(\alpha)} &=& {\sum_{j=1}^c S_j^{(\alpha)}}+2M_1^{(\alpha)}-N_1^{(\alpha)}-N_2^{(\alpha)} \\
\Leftrightarrow\ {\sum_{i=1}^d Q_i^{(\alpha)}} &=& {\sum_{j=1}^c S_j^{(\alpha)}} \qquad{\square}
\end{eqnarray*}
The second linear relation is also satisfied:
\begin{eqnarray*}
 \sum_{a=3}^\delta  N_a^{(\alpha)} + (M_1-S_1) + (-M_1+N_1+N_2+S_1)   &=& \sum_{l=1}^\gamma  M_l^{(\alpha)}\\
 \sum_{a=1}^\delta  N_a^{(\alpha)} &=& \sum_{l=1}^\gamma  M_l^{(\alpha)}  \qquad{\square}
\end{eqnarray*}
\paragraph{Quadratic relations:}
Now lets have a look at the quadratic relations in \eqref{eq_anomcancel}. There are three different cases. The first where only the new $U(1)$ charges are involved, the second where old and new charges get mixed and the third where only the old charges are considered. Lets start with the first, which is obvious since only few changes were made:
$$
1^2 + 1^2 -  ( (-1)^2 +(-1)^2 ) = 1^2- (-1)^2 \qquad \Leftrightarrow \qquad 0=0 \qquad{\square}
$$
For the second one we find:
\begin{eqnarray*}
								 (M_1-N_1) +(M_1-N_2) - (B - 0) &=&  M_1 -( (M_1-S_1) + 0) \\
\Leftrightarrow\   2M_1 - B  &=& N_1+ N_2 + S_1  \\
\Leftrightarrow\   2M_1  - 2M_1 +N_1 + N_2 + S_1   &=& N_1+ N_2 + S_1\\
\Leftrightarrow\   N_1 + N_2 + S_1  &=& N_1+ N_2 + S_1 \qquad{\square}
\end{eqnarray*}
The  last  only involves the old $U(1)$ charges:
\begin{eqnarray}\nonumber
	& & \sum_{l=1}^\gamma  M_l^{(\alpha)} M_l^{(\beta)} - \sum_{a=3}^\delta  N_a^{(\alpha)} N_a^{(\beta)} - \tilde N_1^{(\alpha)}\tilde N_1^{(\beta)}  - \tilde  N_2^{(\alpha)}\tilde N_2^{(\beta)} \\[-0.2cm]\nonumber
	&&=  \sum_{j=2}^c S_j^{(\alpha)} S_j^{(\beta)} + \tilde S_1^{(\alpha)} \tilde S_1^{(\beta)} + \tilde S_2^{(\alpha)} \tilde S_2^{(\beta)}  - \sum_{i=1}^d Q_i^{(\alpha)} Q_i^{(\beta)} - B^{(\alpha)}B^{(\beta)} - 0  \\\nonumber
  \Leftrightarrow & &	 \sum_{a=1}^2  N_a^{(\alpha)} N_a^{(\beta)} - \tilde N_1^{(\alpha)}\tilde N_1^{(\beta)}  - \tilde  N_2^{(\alpha)}\tilde N_2^{(\beta)} \\[-0.2cm]\nonumber 
  &&=- S_1^{(\alpha)} S_1^{(\beta)} + \tilde S_1^{(\alpha)} \tilde S_1^{(\beta)} + \tilde S_2^{(\alpha)} \tilde S_2^{(\beta)}  - B^{(\alpha)}B^{(\beta)}   \\\nonumber
  \Leftrightarrow	& & \sum_{a=1}^2  N_a^{(\alpha)} N_a^{(\beta)} - (M_1^{(\alpha)}-S_1^{(\alpha)})(M_1^{(\beta)}-S_1^{(\beta)})  - (M_1^{(\alpha)}-B^{(\alpha)})(M_1^{(\beta)}-B^{(\beta)}) \\[-0.2cm]\label{eq_plugg in 1}
  &&=-S_1^{(\alpha)} S_1^{(\beta)}  + (M_1^{(\alpha)}-N_1^{(\alpha)})(M_1^{(\beta)}-N_1^{(\beta)})\\[-0.0cm]
  & &\phantom{aaaaaaaaaa} + (M_1^{(\alpha)}-N_2^{(\alpha)})(M_1^{(\beta)}-N_2^{(\beta)})  - B^{(\alpha)}B^{(\beta)} \,,\nonumber
\end{eqnarray}
where we used the initial quadratic relations from \eqref{eq_anomcancel} in
the first step. As an intermediate step, lets evaluate the third and fourth
term of the right hand side  of
\eq{
\label{eq_plugg in 2}
-(M_1^{(\alpha)}-B^{(\alpha)})(M_1^{(\beta)}-&B^{(\beta)}) =\\
 &- M_1^{(\alpha)}M_1^{(\beta)} - B^{(\alpha)}B^{(\beta)} +M_1^{(\alpha)}B^{(\beta)}+B^{(\alpha)}M_1^{(\beta)}
}
which reads
\begin{eqnarray*}
M_1^{(\alpha)}B^{(\beta)} &=& 2M_1^{(\alpha)}M_1^{(\beta)}-M_1^{(\alpha)}N_1^{(\beta)}-M_1^{(\alpha)}N_2^{(\beta)}-M_1^{(\alpha)}S_1^{(\beta)} \,,\\
M_1^{(\beta)}B^{(\alpha)} &=& 2M_1^{(\beta)}M_1^{(\alpha)}-M_1^{(\beta)}N_1^{(\alpha)}-M_1^{(\beta)}N_2^{(\alpha)}-M_1^{(\beta)}S_1^{(\alpha)} 
\end{eqnarray*}
and hence it folllows
\eq{
M_1^{(\alpha)}B^{(\beta)} + M_1^{(\beta)}B^{(\alpha)} = & 
M_1^{(\alpha)}M_1^{(\beta)}\\
& +(M_1^{(\alpha)}-N_1^{(\alpha)})(M_1^{(\beta)}-N_1^{(\beta)})- N_1^{(\alpha)}N_1^{(\beta)}\\
& +(M_1^{(\alpha)}-N_2^{(\alpha)})(M_1^{(\beta)}-N_2^{(\beta)})- N_2^{(\alpha)}N_2^{(\beta)}\\
& +(M_1^{(\alpha)}-S_1^{(\alpha)})(M_1^{(\beta)}-S_1^{(\beta)})- S_1^{(\alpha)}S_1^{(\beta)}\,.\label{eq_plugg in 3}
}
Pluggin \eqref{eq_plugg in 3} back into \eqref{eq_plugg in 2} and \eqref{eq_plugg in 2} back into \eqref{eq_plugg in 1}, we get
\begin{eqnarray}\nonumber
 & & \sum_{a=1}^2  N_a^{(\alpha)} N_a^{(\beta)} - (M_1^{(\alpha)}-S_1^{(\alpha)})(M_1^{(\beta)}-S_1^{(\beta)}) + M_1^{(\alpha)}M_1^{(\beta)}-M_1^{(\alpha)}M_1^{(\beta)}\\\nonumber
 & & - B^{(\alpha)}B^{(\beta)} +(M_1^{(\alpha)}-N_1^{(\alpha)})(M_1^{(\beta)}-N_1^{(\beta)})- N_1^{(\alpha)}N_1^{(\beta)}\\\nonumber
 & & +(M_1^{(\alpha)}-N_2^{(\alpha)})(M_1^{(\beta)}-N_2^{(\beta)})- N_2^{(\alpha)}N_2^{(\beta)}\\\nonumber
 & & +(M_1^{(\alpha)}-S_1^{(\alpha)})(M_1^{(\beta)}-S_1^{(\beta)})- S_1^{(\alpha)}S_1^{(\beta)}\\\nonumber
 &=& -S_1^{(\alpha)} S_1^{(\beta)}  + (M_1^{(\alpha)}-N_1^{(\alpha)})(M_1^{(\beta)}-N_1^{(\beta)})\\\nonumber
 & & + (M_1^{(\alpha)}-N_2^{(\alpha)})(M_1^{(\beta)}-N_2^{(\beta)})  - B^{(\alpha)}B^{(\beta)} \,,\\\nonumber
 &\Leftrightarrow&  0=0 \qquad{\square} \,.
\end{eqnarray}
Since we did not  assume that $M_1=S_1$, the whole calculation is valid for
both cases described in \ref{subsec_construction of dual models}. 
In fact, it
can be shown that one can exchange an arbitrary number of $G$'s with $F$'s, as
long as at most one uncharged Fermi superfield appears.

\clearpage
\bibliography{rev1}
\bibliographystyle{utphys}

\end{document}